\documentclass[11pt, oneside]{article}

\usepackage{latexsym}
\usepackage{amssymb}
\usepackage{amsmath}
\usepackage{amsthm}
\usepackage{graphicx}
\usepackage{fullpage}

\DeclareMathOperator{\arccot}{arccot}
\def\wedge{\mathcal{W}}

\newcommand*\samethanks[1][\value{footnote}]{\footnotemark[#1]}

\begin{document}
\newtheorem{problem}{Problem}
\newtheorem{definition}[problem]{Definition}
\newtheorem{proposition}[problem]{Proposition}
\newtheorem{corollary}[problem]{Corollary}
\newtheorem{remark}[problem]{Remark}
\newtheorem{lemma}[problem]{Lemma}
\newtheorem{algorithm}[problem]{Algorithm}

\title{Minimum-Area Enclosing Triangle with a Fixed Angle}
\author{Prosenjit Bose\thanks{Computational Geometry Lab, School of Computer Science, Carleton University.} \thanks{This research was partially supported by NSERC (Natural Sciences and Engineering
Research Council of Canada).}
\and Jean-Lou De Carufel\samethanks[1] \thanks{This research was partially supported by FQRNT (Fonds qu\'{e}b\'{e}cois de la recherche sur la
nature et les technologies).} }

\maketitle

\begin{abstract}
Given a set $S$ of $n$ points in the plane and a fixed angle $0<\omega<\pi$, 
we show how to find in $O(n \log n)$ time
all triangles of minimum area with one angle $\omega$ 
that enclose $S$.
We prove that in general,
the solution cannot be written without cubic roots.
We also prove an $\Omega(n \log n)$ lower bound for this problem
in the algebraic computation tree model.
If the input is a convex $n$-gon,
our algorithm takes $\Theta(n)$ time.
\end{abstract}

\section{Introduction}
\label{section introduction}

In geometric optimization, 
the goal is often to find an optimal object 
or optimal placement of an object 
subject to a number of geometric constraints.
Examples include finding the smallest circle 
enclosing a point set~\cite{DBLP:journals/siamcomp/Megiddo83a,Welzl91smallestenclosing}
or finding the smallest circle
enclosing at least $k$ points of a point set of $n$ points
($k\leq n$)~\cite{DBLP:journals/comgeo/EfratSZ94,DBLP:journals/ipl/Matousek95}.
In our setting, 
we study the following problem: 
given a set $S$ of $n$ points in the plane, 
find all the triangles of minimum area 
with a fixed angle $\omega$, 
$0<\omega<\pi$, 
that enclose $S$.
When no constraint is put on the angles,
Klee and Laskowski~\cite{DBLP:journals/jal/KleeL85} 
gave an $O(n \log^2 n)$ time algorithm 
for finding the minimum-area enclosing triangle. 
This was later improved to $O(n\log n)$ 
by O'Rourke et al.~\cite{DBLP:journals/jal/ORourkeAMB86}.
Bose et al.~\cite{DBLP:journals/ijcga/BoseMSS11} 
provided optimal algorithms for the setting where 
one wishes to find the minimum-area isosceles triangles. 
The setting we explore here is in between the two. 
We place a restriction on the angle 
but do not insist on the triangle to be isosceles. 
Our solution, 
which we outline below,
uses ideas from the solutions of Klee and Laskowski
and Bose et al.

The five main steps of the algorithm
are presented in Sections~\ref{section overview preliminaries}, 
\ref{section optimal solution given omega wedge}, 
\ref{section walking around omega cloud},
\ref{section optimal solution apex on circular arc}
and \ref{section putting all together}.
Each section outlines one step,
proves the mathematical formulas involved
and gives its time complexity.
As we present in Section~\ref{section optimal solution apex on circular arc},
at some point,
the algorithm needs to calculate the roots of a fourth degree polynomial.
This is unfortunate when it comes to numerical robustness.
However,
in Section~\ref{section complexity solution},
we show that we cannot dodge
such algebraic expressions.
Finally,
we prove an $\Omega(n \log n)$ lower bound for this problem
in the algebraic computation tree model.

\section{Overview and Preliminaries}
\label{section overview preliminaries}

Since the solution to the general problem 
only needs to consider the vertices
of the convex hull of $S$, 
the first step is to compute the convex hull.
In the remainder of the paper,
we assume that the input is a convex $n$-gon
with vertices given in clockwise order. 

\begin{figure}
\centering
\includegraphics[scale=1]{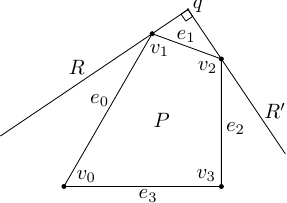}
\caption{In this example,
$P$ is a quadrilateral
and $\omega=\frac{1}{2}\pi$.\label{figure algo-step0}}
\end{figure}

Let $P$ be a convex $n$-gon
(refer to Figure~\ref{figure algo-step0}).
We denote the edges and the vertices of $P$
in clockwise order 
by $e_i$ and $v_i$ for $0\leq i \leq n-1$
(all index manipulation is modulo $n$).
Here and in the following sections,
as we present the algorithm,
we trace each step through the example of Figure~\ref{figure algo-step0}.

We begin with two definitions.
\begin{definition}[$\omega$-wedge]
Let $q$ be a point in the plane
and $\omega$ be an angle 
($0<\omega<\pi$).
Let $R$ and $R'$ be two rays emanating from $q$
such that the angle between $R$ and $R'$ is $\omega$.
We say that the closed set formed by $q$, 
$R$, 
$R'$ 
and the points between $R$ and $R'$
creates an \emph{$\omega$-wedge},
denoted $\wedge(\omega,q,R,R')$.
The point $q$ is called the \emph{apex} of the $\omega$-wedge.
An $\omega$-wedge $W$ \emph{touches} a polygon $P$
when $P\subseteq W$ and both
$R$ and $R'$ touch $P$,
i.e. $R\cap P \neq {\O}$
and $R'\cap P \neq {\O}$
(refer to Figure~\ref{figure algo-step0}).
\end{definition}

For the rest of the paper,
when looking at an $\omega$-wedge facing down,
$R$ represents the left ray and $R'$ represents the right ray
(refer to Figure~\ref{figure algo-step0}).
Also,
when we make the $\omega$-wedge turn around $P$,
we do it clockwise.

\begin{figure}
\centering
\includegraphics[scale=1]{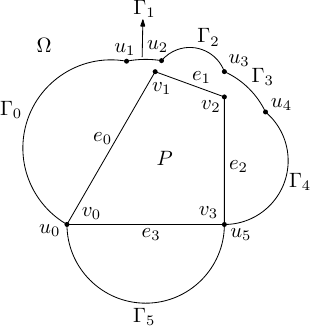}
\caption{{\bf Step 1}: $\Omega$ is the $\frac{1}{2}\pi$-cloud of $P$ ($n'=6$).\label{figure algo-step1}}
\end{figure}

\begin{definition}[$\omega$-cloud]
Let $P$ be a convex $n$-gon
and $\omega$ be an angle 
($0<\omega<\pi$).
By rotating an $\omega$-wedge around $P$
while continually touching $P$,
the apex traces a sequence of circular arcs
that we call an \emph{$\omega$-cloud}
(refer to Figure~\ref{figure algo-step1}).
\end{definition}

There are many technical details involved 
in the solution to this problem.
Before getting too caught up in these details,
let us first review the general approach to our solution.

Since we only consider enclosing triangles 
with an angle of $\omega$, 
each optimal triangle can be constructed 
from an $\omega$-wedge that touches $P$.
Therefore, 
we consider all possible $\omega$-wedges that touch $P$.
The apices of these $\omega$-wedges
lie on an $\omega$-cloud $\Omega$ which consists 
of a linear number of pieces of circular arcs 
(refer to~\cite{DBLP:journals/ijcga/BoseMSS11}).
Then, 
for each of these $\omega$-wedges,
it is possible to find the minimal triangle
by identifying a third side.
For the triangle to be optimal,
the midpoint of this third side
has to touch $P$
(see Proposition~\ref{proposition n-gon wedge minimal triangle} 
and Corollary~\ref{corollary n-gon wedge minimal triangle}).
Hence,
for each $\omega$-wedge touching $P$,
there is one and only one triangle to consider for optimality.

Moreover,
when the apex $q$ of an $\omega$-wedge $W$
moves clockwise along the $\omega$-cloud $\Omega$,
the midpoint $m$ of the third side of the optimal triangle
moves clockwise along $P$
(see Lemma~\ref{lemma q turns clockwise and so does m}).
As $W$ moves,
we note the positions of $q$
where $m$ leaves an edge of $P$
and where $m$ enters a new edge of $P$.
These positions of $q$ are important event points.
They divide $\Omega$ into a linear number of components
(see Section~\ref{section walking around omega cloud}).
Let $\mathcal{C}$ be one of these components.
We prove that the minimum-area triangle having a vertex 
(hence an angle $\omega$) on $\mathcal{C}$
can be computed in constant time
(see Lemmas~\ref{lemma arc fixed point minimal triangle},
\ref{lemma arc fixed line minimal triangle},
\ref{lemma vertex fixed point minimal triangle}
and~\ref{lemma vertex fixed line minimal triangle}).
We then have a linear number of candidates
(one for each piece of $\Omega$)
to consider rather than infinitely many
if we had to take all possible 
$\omega$-wedges touching $P$ into account.
What remains is to identify
the optimal ones
from these linear candidates.
With this in mind,
the technical details will fall into place.

\begin{description}
\item[Step 1] Compute the $\omega$-cloud around $P$
and denote it by $\Omega$.
\end{description}
The $\omega$-cloud $\Omega$ consists of $n'=O(n)$ circular arcs
that we denote in clockwise order by $\Gamma_i$ for $0\leq i \leq n'-1$.
The intersection point of $\Gamma_i$ and $\Gamma_{i+1}$ 
is denoted by $u_{i+1}$ for $0\leq i \leq n'-1$
(refer to Figure~\ref{figure algo-step1}).
We also refer to $\Gamma_i$ by the closed set $[u_i,u_{i+1}]$
containing all the points of $\Gamma_i$.
{\bf Step 1} takes $O(n)$ time
(refer to~\cite{DBLP:journals/ijcga/BoseMSS11}).

\section{Optimal Solution Given a Fixed $\omega$-Wedge}
\label{section optimal solution given omega wedge}

In this section,
we present a routine that computes the minimum-area enclosing triangle
from a fixed $\omega$-wedge touching $P$
(see Algorithm~\ref{algorithm minimum enclosing triangle fixed wedge}).
Then we describe {\bf Step 2}
that uses this routine.

Take any $\omega$-wedge $W=\wedge(\omega,q,R,R')$ 
that touches $P$.
Find $b\in R$
(respectively $c\in R'$)
such that $P$ is enclosed in $\triangle qbc$
and the midpoint $m$ of $bc$ is on $P$.
We claim that
\begin{enumerate}
\item it is always possible to find $b$ and $c$ satisfying these properties,

\item $\triangle qbc$ is the minimum-area triangle enclosing $P$
that can be constructed with $\wedge(\omega,q,R,R')$,
 
\item it takes $O(n)$ time to compute $\triangle qbc$.
\end{enumerate}
These claims are proven 
in Proposition~\ref{proposition n-gon wedge minimal triangle},
Algorithm~\ref{algorithm minimum enclosing triangle fixed wedge},
Lemma~\ref{lemma fixed wedge minimal triangle}
and Corollary~\ref{corollary n-gon wedge minimal triangle}.

We first address the case where $q$ is outside of $P$.
\begin{proposition}
\label{proposition n-gon wedge minimal triangle}
Let $P$ be a convex $n$-gon
and $q$ be a point outside of $P$.
Let $\triangle qbc$ be the triangle that encloses $P$ 
such that the midpoint $m$ of $bc$ lies on $P$.
The following is true:
\begin{enumerate}
\item\label{proposition n-gon wedge minimal triangle item exists unique}
$\triangle qbc$ exists and is unique
(hence it is well-defined).

\item\label{proposition n-gon wedge minimal triangle item minimality}
Among all triangles that enclose $P$
and have $q$ as a vertex,
$\triangle qbc$ is the unique one of minimum area.
\end{enumerate}
\end{proposition}

Proposition~\ref{proposition n-gon wedge minimal triangle}
was first proven by Klee and Laskowski
(refer to~\cite[Lemma~1.2]{DBLP:journals/jal/KleeL85}).
We propose a slightly different proof.
Among other things,
we stress the fact that $\triangle qbc$ always exists.
\proof\
\begin{enumerate}
\item Take $b$ and $c$ on the rays that support $P$ from $q$
such that an edge $e$ of $P$ is flush with $bc$
(refer to Figure~\ref{figure KleeLaskowski-1}).
\begin{figure}
\centering
\includegraphics[scale=1]{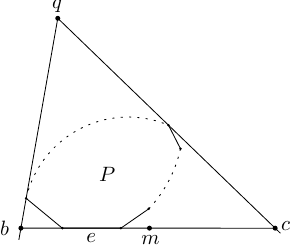}
\caption{$\triangle qbc$ exists.\label{figure KleeLaskowski-1}}
\end{figure}
If $m\in e$, 
then we are done.
If not,
suppose without loss of generality
that $m$ is between $e$ and $c$.
Move $b$ along its ray such that
$b$ gets farther from $q$ 
and $bc$ stays tangent to $P$.
Therefore $c$ gets closer to $q$.
Since $m$ moves continuously
as $b$ moves along its ray,
$m$ will eventually touch an edge of $P$.
This continuity argument implies the existence of $\triangle qbc$.

Suppose there are two triangles,
namely $\triangle qbc$ and $\triangle qb'c'$
with midpoints $m$ and $m'$ respectively.
We show that these two triangles are equal.
There are three cases to consider:
(1) $m = m'$,
(2) $m \neq m'$ and both belong to the same edge of $P$
or (3) $m \neq m'$ and both belong to different edges of $P$.

\begin{enumerate}
\item[(1)]
If $m = m'$ is not on a vertex of $P$,
then it lies in the interior of an edge of $P$
(refer to Figure~\ref{figure KleeLaskowski-2}).
\begin{figure}
\centering
\includegraphics[scale=1]{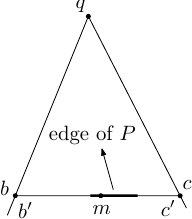}
\caption{$m=m'$ is not a vertex of $P$.\label{figure KleeLaskowski-2}}
\end{figure}
Hence this aforementioned edge is in $bc \cap b'c'$.
This implies that $b=b'$ and $c=c'$.

Suppose $m = m'$ is on a vertex of $P$.
Triangles $\triangle mbb'$ and $\triangle mcc'$ are congruent by the following:
(refer to Figure~\ref{figure KleeLaskowski-3}).
\begin{figure}
\centering
\includegraphics[scale=1]{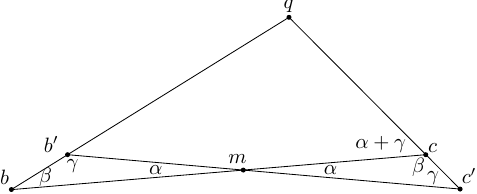}
\caption{$m=m'$ is a vertex of $P$.\label{figure KleeLaskowski-3}}
\end{figure}
\begin{itemize}
\item $|b'm| = |c'm|$ since $m$ is the midpoint of $b'c'$.

\item $\angle b'mb = \angle c'mc$ since they are vertical angles.

\item $|bm| = |cm|$ since $m$ is the midpoint of $bc$.
\end{itemize}
Therefore $\angle mbb' = \angle mcc'$
and $\angle mb'b = \angle mc'c$.
Hence we have the following:
\begin{eqnarray*}
\alpha &=& \angle b'mb = \angle c'mc \enspace,\\
\beta &=& \angle mbb' = \angle mcc' \enspace,\\
\gamma &=& \angle mb'b = \angle mc'c \enspace,
\end{eqnarray*}
where $\alpha+\beta+\gamma = \pi$.
So $\angle bcq = \alpha + \gamma$
because $\angle bcq$ and $\angle c'cb$ are supplementaries.
Therefore 
$\angle bqc = \pi - \alpha - \beta - \gamma = 0$,
which is impossible unless $\alpha = 0$.
We conclude that $b=b'$ and $c=c'$.

\item[(2)] Suppose that $m \neq m'$ and they belong to the same edge of $P$.
If neither $m$ nor $m'$ is a vertex,
then this situation is similar to the one in Figure~\ref{figure KleeLaskowski-2},
hence $b=b'$, $c=c'$ and $m=m'$.
This is a contradiction so this situation is impossible.

Without loss of generality,
suppose that $m'$ is a vertex of $P$
and that $|m'b| < |m'c|$.
Let $b''$ be the point on the line through $b'c'$
such that the line segments $bb''$ and $cc'$ are parallel
(refer to Figure~\ref{figure KleeLaskowski-4}).
\begin{figure}
\centering
\includegraphics[scale=1]{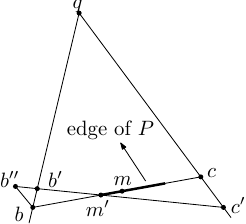}
\caption{$m'$ is a vertex of $P$.\label{figure KleeLaskowski-4}}
\end{figure}
Triangles $\triangle m'bb''$ and $\triangle m'cc'$ are similar by the following:
\begin{itemize}
\item $\angle bm'b'' = \angle cm'c'$ since they are vertical angles.

\item $\angle m'bb'' = \angle m'cc'$ since they are alternate angles.
\end{itemize}
Therefore,
since $|m'b| < |m'c|$,
then $|m'b''| < |m'c'|$.
However,
since $m'$ is the midpoint of $b'c'$,
$|m'b''| > |m'c'|$.
This is a contradiction so this situation is impossible.

\item[(3)] Suppose $m \neq m'$ do not belong to the same edge of $P$.
This case is similar to the case where
$m \neq m'$ and both points 
are on the same edge of $P$.
\end{enumerate}

\item Let $\triangle qb'c'$ be a triangle 
of minimum area that circumbscribes $P$
and has $q$ as a vertex
(refer to Figure~\ref{figure KleeLaskowski-5}).
\begin{figure}
\centering
\includegraphics[scale=1]{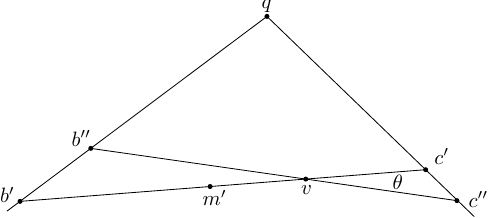}
\caption{$\triangle qbc$ has minimum area.\label{figure KleeLaskowski-5}}
\end{figure}
Suppose for a contradiction
that the midpoint $m'$ of $b'c'$
does not lie on one of the edges of $P$
(refer to Figure~\ref{figure KleeLaskowski-5}).

Since $P$ is convex
and $\triangle qb'c'$ circumbscribes $P$,
the following construction is possible.
Let $v \in b'c'$ be the vertex of $P$ closest to $m'$.
Without loss of generality, 
suppose $vc' < vb'$.
Let $b''c''$ be a line segment that goes through $v$
and such that $b'' \in qb'$
and $c''$ lies on the line through $q$ and $c'$.
Let $\theta = \angle c'vc''$.
We have $\angle b'vb'' = \angle c'vc''$ since they are vertical angles.
If $\theta$ is sufficently small,
then $vc'' < vb''$
and $\triangle qb''c''$ circumbscribes $P$.

Therefore
\begin{eqnarray*}
area(\triangle qb'c') &=& area(\triangle b'vb'') + area([qb''vc']) \\
&=& \frac{1}{2}|vb'||vb''|\sin(\theta) + area([qb''vc']) \\
&>& \frac{1}{2}|vc'||vc''|\sin(\theta) + area([qb''vc']) \\
&=& area(\triangle c'vc'') + area([qb''vc']) \\
&=& area(\triangle qb''c'') \enspace,
\end{eqnarray*}
which contradicts the fact that $\triangle qb'c'$ has minimum area.
So $m'$ lies on one of the edges of $P$
and $\triangle qb'c'$ is a local minimum
among triangles circumbscribing $P$
and having $q$ as a vertex.
We proved in Point~\ref{proposition n-gon wedge minimal triangle item exists unique}
that there exists one and only one such triangle.
\qed
\end{enumerate}

Given the setting of Proposition~\ref{proposition n-gon wedge minimal triangle},
we show how to compute $\triangle qbc$ in $O(n)$ time.
In Algorithm~\ref{algorithm minimum enclosing triangle fixed wedge},
the edges of $P$ are considered in clockwise order.
The variable $edge$ indicates
the edge of $P$
that is currently being considered
and $\textrm{next}(edge)$ represents the edge next to it in clockwise order.
Algorithm~\ref{algorithm minimum enclosing triangle fixed wedge} uses
Lemma~\ref{lemma fixed wedge minimal triangle}
that is presented
after this routine.
\begin{algorithm}(Minimum-Area Enclosing Triangle with a Fixed $\omega$-Wedge)
\label{algorithm minimum enclosing triangle fixed wedge}
\begin{itemize}
\item INPUT: A convex $n$-gon $P$ 
and an $\omega$-wedge $W=\wedge(\omega,q,R,R')$ touching $P$
which supports $P$ at vertices $v_0$ and $v_k$
(refer to Figure~\ref{figure KleeLaskowski-6}).
\begin{figure}
\centering
\includegraphics[scale=1]{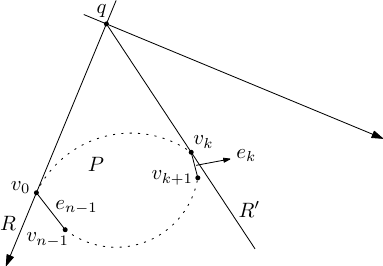}
\caption{Computing $\triangle qbc$ takes $O(n)$ time.\label{figure KleeLaskowski-6}}
\end{figure}

\item OUTPUT: The minimum-area triangle enclosing $P$
that can be constructed with $\wedge(\omega,q,R,$ $R')$.
\end{itemize}
\begin{enumerate}
\item $edge \longleftarrow e_k$.

\item\label{step edge intersects wedge} 
If the line through $edge$ does not intersect $R$
or $R'$,
\begin{itemize}
\item $edge \longleftarrow \textrm{next}(edge)$.

\item Go to~\ref{step edge intersects wedge}.
\end{itemize}

\item\label{step midpoint on P}
Let $b$ (respectively $c$)
be the intersection point
of $R$ (respectively $R'$)
and the line through $edge$.
If the midpoint of $bc$
is on $edge$, 
return $\triangle qbc$.

\item If the midpoint of $bc$ is between $b$ and $edge$,
\begin{itemize}
\item $edge \longleftarrow \textrm{next}(edge)$.

\item Go to~\ref{step midpoint on P}.
\end{itemize}

\item If the computation reaches this step,
it means that the midpoint of $bc$ is between $edge$ and $c$.
\begin{itemize}
\item Place the Cartesian coordinate system on $W$
such that $q = (0,0)$
and $R$ is the positive $x$-axis
(refer to Figure~\ref{figure KleeLaskowski-6}).

\item Let $v=(s,t):=edge \cap \textrm{previous}(edge)$.

\item Take $b=(2(s-t\cot(\omega)),0)$ and $c=(2t\cot(\omega),2t)$
(refer to the proof of Lemma~\ref{lemma fixed wedge minimal triangle}).

\item Return $\triangle qbc$.
\end{itemize}
\end{enumerate}
\end{algorithm}
In the worst case,
Algorithm~\ref{algorithm minimum enclosing triangle fixed wedge} considers all the edges of $P$.
Since it spends $O(1)$ time per edge,
it takes $O(n)$ time total.

\begin{lemma}
\label{lemma fixed wedge minimal triangle}
Let $R$ and $R'$ be two lines
intersecting at $q$
and let $\omega$ ($0<\omega<\pi$)
be the angle between $R$ and $R'$.
Let $v$ be a point in $\wedge(\omega,q,R,R')$
($v\not\in R$ and $v\not\in R'$).
In $O(1)$ time,
we can compute $b\in R$
and $c\in R'$ such that
$b$, $v$ and $c$ lie on a single line
and $v$ is the midpoint of $bc$.
\end{lemma}

\begin{figure}
\centering
\includegraphics[scale=1]{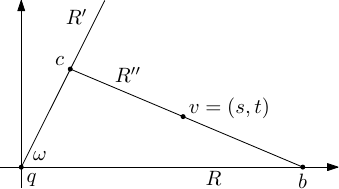}
\caption{Proof of Lemma~\ref{lemma fixed wedge minimal triangle}.\label{fig alpha-wedge}}
\end{figure}

\proof
Without loss of generality,
suppose $R$ is the $x$-axis,
$R'$ is the line\footnote{
In what follows,
several algebraic expressions are written using $\cot(\cdot)$
where it seems that they could be simplified by using $\tan(\cdot)$ instead.
However,
we write $\cot(\cdot)$ in order to 
properly deal with the angle $\frac{1}{2}\pi$.} 
$x-\cot(\omega)y=0$
and let $v = (s,t)$
(refer to Figure~\ref{fig alpha-wedge}).
We have $t>0$ and $s-\cot(\omega)t\neq 0$
since $v\not\in R$ and $v\not\in R'$,
respectively.

Let $R''$ be a line containing $v$ such that
$R$ is not parallel to $R''$
and $R'$ is not parallel to $R''$.
The line $R''$ is the one containing the points
$b$ and $c$ we are looking for.
There are two cases to consider:
(1) $R''$ is not vertical
or (2) $R''$ is vertical.

\begin{enumerate}
\item[(1)] Suppose $R''$ is not vertical.
Let $b$ be the intersection point of $R$ and $R''$,
and $c$ the intersection point of $R'$ and $R''$.
The general equation of $R''$ is $y = \lambda (x-s)+t$.
Note that $\lambda\neq 0$ since $R$ is not parallel to $R''$
and $1-\lambda\cot(\omega) \neq 0$ since $R'$ is not parallel to $R''$.
Calculating the intersection point of $R$ and $R''$,
and of $R'$ and $R''$,
we find that the general coordinates of $b$ and $c$ are
$b=\left(\frac{\lambda s-t}{\lambda},0\right)$
and $c = \left(\frac{t-\lambda s}{1-\lambda\cot(\omega)}\cot(\omega),\frac{t-\lambda s}{1-\lambda\cot(\omega)}\right)$.
We are looking for $\lambda$ such that $|bv| = |vc|$,
so we need to isolate $\lambda$ in
$$\left(\frac{\lambda s-t}{\lambda}-s\right)^2+\left(0-t\right)^2 = \left(\frac{t-\lambda s}{1-\lambda\cot(\omega)}\cot(\omega)-s\right)^2+\left(\frac{t-\lambda s}{1-\lambda\cot(\omega)}-t\right)^2 \enspace,$$
which solves to $\lambda = \frac{t}{2t\cot(\omega)-s}$.
Therefore
$b=(2(s-t\cot(\omega)),0)$ and $c=(2t\cot(\omega),2t)$.

\item[(2)] Suppose $R''$ is vertical.
Therefore 
$0<\omega<\frac{1}{2}\pi$,
otherwise $R''$ is not the expected line.
Using the notation of the previous case,
we get that the equation of $R''$ is $x=s$.
Moreover,
$b=(s,0)$ and $c=(s,s\tan(\omega))$.
We want $|bv| = |vc|$,
which means
$$(s-s)^2+(0-t)^2 = (s-s)^2+(s\tan(\omega)-t)^2 \enspace.$$ 
So this situation occurs 
if and only if $2t=s\tan(\omega)$
and the solution is $b=(s,0)=(2(s-t\cot(\omega)),0)$ 
and $c=(s,s\tan(\omega))=(2t\cot(\omega),2t)$.
\end{enumerate}

Hence the global solution is
$b=(2(s-t\cot(\omega)),0)$ 
and $c=(2t\cot(\omega),2t)$
in all cases.
\qed

Note that what we obtained
is more general than our claims.
Denote by $T$ the triangle
computed by Proposition~\ref{proposition n-gon wedge minimal triangle}
together with Algorithm~\ref{algorithm minimum enclosing triangle fixed wedge}
and Lemma~\ref{lemma fixed wedge minimal triangle}.
Because $q$ is outside of $P$
and $\wedge(\omega,q,R,R')$ touches $P$,
not only is $T$ the minimum-area triangle
enclosing $P$ 
that can be constructed with $\wedge(\omega,q,R,R')$,
it is also the minimum-area triangle
enclosing $P$
and having $q$ as a vertex.

When $q$ is on $P$,
we can compute
the minimum-area triangle
enclosing $P$ 
that can be constructed with $\wedge(\omega,q,R,R')$ in the same way.
However,
in general,
it does not correspond to the minimum-area triangle
enclosing $P$
and having $q$ as a vertex
(see Figure~\ref{fig algo-step2-SpecialCase}
and the discussion at the beginning of Section~\ref{section walking around omega cloud}).
\begin{corollary}
\label{corollary n-gon wedge minimal triangle}
Consider an $\omega$-wedge $W=\wedge(\omega,q,R,R')$
touching $P$
such that $q$ is on $P$.
Then $q$ is on one of the vertices of $P$
and the minimum-area triangle enclosing $P$ 
that can be constructed with $W$
can be computed in $O(n)$ time. 
\end{corollary}

Therefore,
given any fixed $\omega$-wedge $W$ touching $P$,
we can compute the minimum-area triangle enclosing $P$
that can be constructed with $W$.
The midpoint $m$ of the third side 
of this optimal triangle
has to be on $P$.
In {\bf Step 3}
(see Section~\ref{section walking around omega cloud}),
we need to know the exact position of $m$ 
for one fixed $\omega$-wedge touching $P$.
However,
it does not matter
for what $\omega$-wedge touching $P$
we know the position of $m$,
as long as we know it for one.
Therefore, 
in {\bf Step 2},
we fix an arbitrary $\omega$-wedge $W$ touching $P$
and we compute the position of $m$.

\begin{description}
\item[Step 2] Let $q \in \Gamma_0$ be such that $q=u_0$.
Consider the $\omega$-wedge $W=\wedge(\omega,q,R,R')$ 
that touches $P$
(refer to Figure~\ref{figure algo-step2}).
\begin{figure}
\centering
\includegraphics[scale=1]{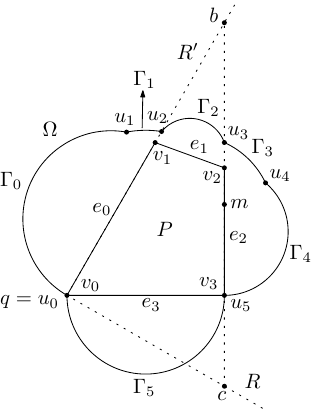}
\caption{{\bf Step 2}: 
$\triangle qbc$ is constructed with $\wedge(\frac{1}{2}\pi,q,R,R')$
where $q=u_0\in\Gamma_0$.\label{figure algo-step2}}
\end{figure}
Apply Algorithm~\ref{algorithm minimum enclosing triangle fixed wedge}
with $P$ and $W$.
\end{description}

Consider Figure~\ref{figure algo-step1}.
Any $\omega$-wedge touching $P$ 
and having its vertex in $\Gamma_0\setminus\{u_0,u_1\}$
is such that 
$v_0\in R'$
and $v_1\in R$.
This property easily generalizes to the case 
where the apex of the $\omega$-wedge is on $u_0$ or on $u_1$.
Therefore,
when we specify $q\in \Gamma_0$ and $q=u_0$,
we imply that
$v_0\in R'$
and $v_1\in R$,
and hence the $\omega$-wedge is unique.
However,
in the example of Figure~\ref{figure algo-step2},
there are infinitely many other $\omega$-wedges touching $P$
and having $q=u_0$ as a vertex to consider
because $u_0$ is on a vertex of $P$
(indeed $u_0=v_0$).
The algorithm will consider these other $\omega$-wedges later
(see Section~\ref{section walking around omega cloud}).

From Proposition~\ref{proposition n-gon wedge minimal triangle},
Algorithm~\ref{algorithm minimum enclosing triangle fixed wedge},
Lemma~\ref{lemma fixed wedge minimal triangle}
and Corollary~\ref{corollary n-gon wedge minimal triangle},
{\bf Step 2} takes $O(n)$ time.

\section{Walking Around the $\omega$-Cloud}
\label{section walking around omega cloud}

In the previous section,
we computed the minimum-area triangle
enclosing $P$
that can be constructed with a given $\omega$-wedge
$\wedge(\omega,q,R,R')$.
The next step is to consider
all possible $\omega$-wedges touching $P$
and having their apex on $\Omega$.
For each of these $\omega$-wedges,
we can compute the minimum-area triangle enclosing $P$.
The final solution is the minimum among all these triangles.
Of course,
there are infinitely many such triangles.
However,
we show that we need to consider 
only $O(n)$ of these triangles.

\begin{figure}
\centering
\includegraphics[scale=1]{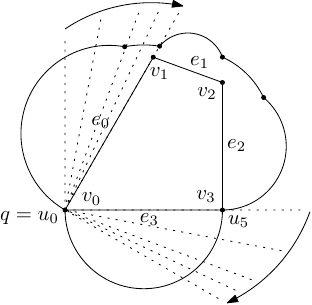}
\caption{$\frac{1}{2}\pi$-wedge turning clockwise around a corner.\label{fig algo-step2-SpecialCase}}
\end{figure}

In what follows,
we consider the points of $\Omega$ 
and the related $\omega$-wedges
in clockwise order.
When thinking of all possible $\omega$-wedges touching $P$
and having their apex on $\Omega$,
one must pay attention to the following case.
If $q$ is the endpoint of one of the circular arcs $u_i$ of $\Omega$,
and if at the same time this endpoint is one of the vertices $v_j$ of $P$,
then there are infinitely many $\omega$-wedges to consider
(refer to Figure~\ref{fig algo-step2-SpecialCase}).
This situation occurs when $\omega$ is greater 
than the angle between the two edges $e_{j-1}$ and $e_j$.
Then the $\omega$-wedge turns
around the corner created by $e_{j-1}$ and $e_j$
while its apex $q$ stays on the vertex $v_j$.

The following lemma
describes the behaviour of $m$
(the midpoint of $bc$ constructed in {\bf Step 2})
as $q$ moves clockwise along $\Omega$.
Some hypotheses of this lemma
are written using the expression ``close enough''.
Its meaning is twofold.
First,
as an $\omega$-wedge $W=\wedge(\omega,q,R,R')$ touching $P$
moves continuously
such that its apex $q$ stays on $\Omega$,
the midpoint $m$ of $bc$ 
also moves continuously.
Secondly,
let $e$ be an edge of $P$.
If $m\in e$,
it is possible to move $q$ 
such that $m$ stays on $e$.
Such a displacement of $q$ has to be 
of less than $\varepsilon$ for an $\varepsilon>0$
and such an $\varepsilon$ always exists.

What needs to be shown
is that $m$ moves clockwise along $P$.
\begin{lemma}
\label{lemma q turns clockwise and so does m}
As an $\omega$-wedge $W=\wedge(\omega,q,R,R')$ touching $P$ moves clockwise
such that its apex $q$ stays on $\Omega$,
the midpoint $m$ of $bc$ 
---the third edge of the minimum-area triangle enclosing $P$
that can be constructed with $W$---
moves clockwise along $P$.

Specifically,
take $q',q'' \in \Omega$
with $q''$ clockwise from $q'$.
Denote by $m'$
(respectively by $m''$)
the midpoint of the third side of the minimum-area triangle
associated with the $\omega$-wedge touching $P$ with apex $q'$
(respectively with apex $q''$).
\begin{enumerate}
\item\label{lemma q turns clockwise and so does m item q' neq q''} 
If $q'\in \Gamma_i\setminus\{u_{i+1}\}$ ($0\leq i \leq n'-1$),
$q''\in\,]q',u_{i+1}]$,
and $q'$ and $q''$ are close enough to each other,
then we have the following:
\begin{enumerate}
\item\label{lemma q turns clockwise and so does m item q' neq q'' item m' neq m''}  
If $m'\in e_j \setminus \{v_j,v_{j+1}\}$ ($0\leq j \leq n-1$),
then $m'' \in\, ]m',v_{j+1}]\subset e_j$.

\item If $m'=v_j$ ($0\leq j \leq n-1$),
then $m''\in e_j$
(possibly $m''=m'$).
\end{enumerate}

\item\label{lemma q turns clockwise and so does m item q' = q'' = ui}
Suppose $q'=q''=u_i$ ($0\leq i \leq n'-1$)
and $u_i=v_{\ell}$ ($0\leq \ell \leq n-1$).
Let $W'$ and $W''$ be two different $\omega$-wedges
touching $P$ with $q'=q''=u_i$ as an apex.
If $W''$ is clockwise from $W'$
(refer to Figure~\ref{fig algo-step2-SpecialCase})
and $W'$ and $W''$ are close enough to each other,
then we have the following:
\begin{enumerate}
\item If $m'\in e_j \setminus \{v_j,v_{j+1}\}$ ($0\leq j \leq n-1$),
then $m'' \in\, ]m',v_{j+1}]\subset e_j$.

\item If $m'=v_j$ ($0\leq j \leq n-1$),
then $m''\in e_j$
(possibly $m''=m'$).
\end{enumerate} 
\end{enumerate}
\end{lemma}

\begin{figure}
\centering
\includegraphics[scale=1]{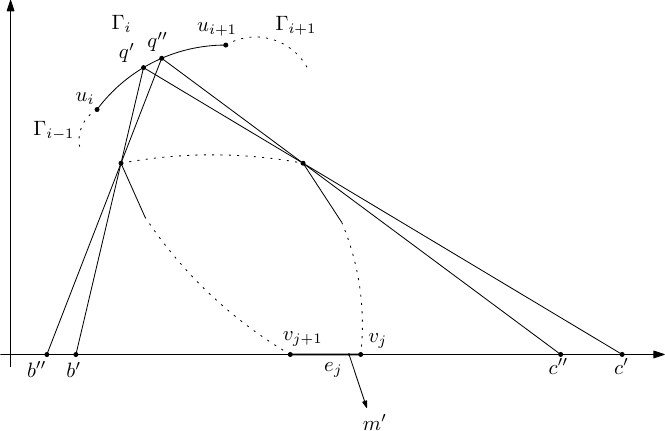}
\caption{Proof of Lemma~\ref{lemma q turns clockwise and so does m}.\label{figure clockwise}}
\end{figure}

\proof Without loss of generality,
$b'c'$ is on the $x$-axis
(refer to Figure~\ref{figure clockwise}).
\begin{enumerate}
\item
\begin{enumerate}
\item
If $q'$ and $q''$ are close enough to each other,
then $m'' \in e_j$.
Therefore,
$b''=(b''_x,0)$,
$b'=(b'_x,0)$,
$c''=(c''_x,0)$
and $c'=(c'_x,0)$,
for $b'_x,b''_x,c'_x,c''_x \in \mathbb{R}$
such that $b''_x < b'_x < c''_x < c'_x$.
Hence,
$m''=\frac{b''_x+c''_x}{2}<\frac{b'_x+c'_x}{2}=m'$,
so $m'' \in\, ]m',v_{j+1}]\subset e_j$.

\item
If $q'$ and $q''$ are close enough to each other,
then the only situation to discard is $m''\in e_{j-1}\setminus\{v_j\}$.
Thus,
suppose $m''\in e_{j-1}\setminus\{v_j\}$ for a contradiction.
Therefore,
$m'$ and $m''$ both belong to $e_{j-1}$.
Hence,
an argument similar to the one of the previous case leads to $m''=v_j$, 
which is a contradiction.
\end{enumerate}

\item The proof is similar to the one of Point \ref{lemma q turns clockwise and so does m item q' neq q''}.
\qed
\end{enumerate}

Therefore,
the midpoint $m$ of the third side of the triangle
is either a vertex of $P$
or on an edge of $P$.
The goal of {\bf Step 3} is to identify 
the sections of $\Omega$
where the midpoint is a vertex
and the sections of $\Omega$
where the midpoint is on an edge of $P$.

\begin{description}
\item[Step 3] Move the apex $q$ of the $\omega$-wedge
clockwise along $\Omega$.
Maintain $b$, $c$ and $m$
as defined in {\bf Step 2}
(refer to Section~\ref{section optimal solution given omega wedge}).
Collect all of the following three types of event points
(refer to Figure~\ref{figure algo-step3}).
\begin{figure}
\centering
\includegraphics[scale=1]{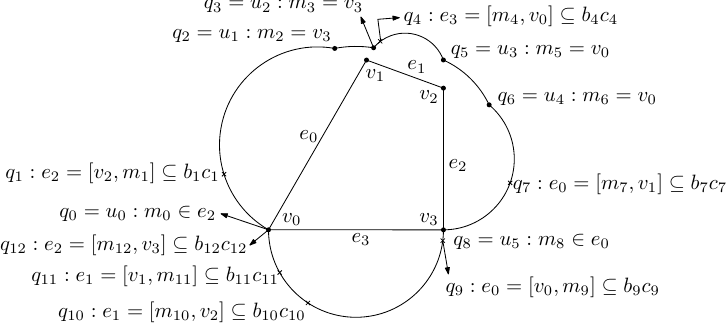}
\caption{{\bf Step 3}:
$q_0=u_0=v_0$ is such that $m_0\in e_2$.
Note that $u_0=v_0$ gives birth to two different event points,
namely $q_0$ and $q_{12}$.
They correspond to the $\frac{1}{2}\pi$-wedge touching $P$
and having $u_0\in\Gamma_0$ as a vertex,
and the $\frac{1}{2}\pi$-wedge touching $P$
and having $u_0\in\Gamma_5$ as a vertex.
$q_2 = u_1:m_2=v_3$ means that $q_2=u_1$,
$m_2=v_3$
and $b_2c_2 \cap P = \{v_3\}$.
It is an event point of the first type.
At such a place, 
$m$ does not move even though $q$ does.
$q_4:e_3=[m_4,v_0]\subseteq b_4c_4$ means that
$m_4=v_3$ and $e_3\subseteq b_4c_4$.
It is an event point of the second type.
$q_9:e_0=[v_0,m_9]\subseteq b_9c_9$ means that
$m_9=v_1$ and $e_0\subseteq b_9c_9$.
It is an event point of the third type.\label{figure algo-step3}}
\end{figure}
\begin{description}
\item[Type 1]
$q$ is on the intersection point of two consecutive circular arcs of $\Omega$ 
for an $i$ with $0\leq i \leq n'-1$.
Formally,
$q=u_i$ for an $i$ with $0\leq i \leq n'-1$.

\item[Type 2]
$q$ is such that the third side $bc$ of the triangle
is on an edge $e_i$ of $P$
(for an $i$ with $0\leq i \leq n-1$)
and the midpoint $m$ of $bc$
is on the first vertex $v_i$ of $e_i$
(when the vertices of $P$ are considered in clockwise order).
Formally,
$q$ is such that $m=v_i$ and $e_i=[v_i,v_{i+1}]=[m,v_{i+1}] \subseteq bc$ 
for an $i$ with $0\leq i \leq n-1$.

\item[Type 3]
$q$ is such that the third side $bc$ of the triangle
is on an edge $e_i$ of $P$
(for an $i$ with $0\leq i \leq n-1$)
and the midpoint $m$ of $bc$
is on the last vertex $v_{i+1}$ of $e_i$
(when the vertices of $P$ are considered in clockwise order).
Formally,
$q$ is such that $m=v_{i+1}$ and $e_i=[v_i,v_{i+1}]=[v_i,m] \subseteq bc$ 
for an $i$ with $0\leq i \leq n-1$.
\end{description}

For each event point $q_i$,
save its type together with the location of $m_i$,
the midpoint of the third side of the triangle.
\end{description}

\begin{figure}
\centering
\includegraphics[scale=1]{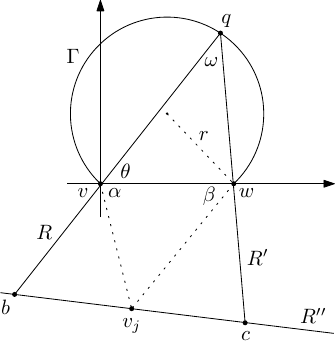}
\caption{Proof of Lemma~\ref{lemma arc fixed midpoint on a line}.\label{arc-circle-segment-point-proof}}
\end{figure}

It is easy to find the event points of the first type
and the following two lemmas
show how to find the event points 
of the second and third type.
Lemma~\ref{lemma arc fixed midpoint on a line}
helps identifying event points related
to Case~\ref{lemma q turns clockwise and so does m item q' neq q''}
of Lemma~\ref{lemma q turns clockwise and so does m}.
As for Lemma~\ref{lemma vertex fixed midpoint on a segment},
it helps identifying event points related
to Case~\ref{lemma q turns clockwise and so does m item q' = q'' = ui}
of Lemma~\ref{lemma q turns clockwise and so does m}.
\begin{lemma}
\label{lemma arc fixed midpoint on a line}
Let $\Gamma = [v,w]$ be an arc of a circle,
$R''$ be a line
and $v_j \in R''$ be a point.
It is possible to find a triangle $\triangle qbc$ 
such that $q \in \Gamma$,
$b,c\in R''$
and $v_j$ is the midpoint of $bc$
(or decide that there is no such point)
in $O(1)$ time
(refer to Figure~\ref{arc-circle-segment-point-proof}).
\end{lemma}

\proof
Without loss of generality,
$\Gamma$ is the locus of the point $q$
such that $\angle vqw = \omega$.
Hence we can take
$v=(0,0)$,
$v_j=(s,t)$ with $t > 0$
and $w=(2r\sin(\omega),0)$,
where $r$ is the radius of $\Gamma$.
Let $\alpha = \angle v_jvw$,
$\beta = \angle v_jwv$
and $q$ be any point of $\Gamma$
(refer to Figure~\ref{arc-circle-segment-point-proof}).
There are three cases to consider:
(1) either $R''$ has strictly negative slope,
(2) $R''$ has non-negative slope
(3) or $R''$ is vertical.

\begin{enumerate}
\item[(1)] Suppose $R''$ has strictly negative slope.
The general equation for $R''$ is $y=\lambda x+\mu$ ($\lambda<0$)
and hence $v_j=(s,\lambda s+\mu)$.
Note $\theta = \angle qvw$.
Also let $R$ 
(respectively $R'$)
be the line through $q$ and $v$
(respectively through $q$ and $w$).
Let $b$ 
(respectively $c$)
be the intersection point of $R$ and $R''$
(respectively of $R'$ and $R''$).

Trivially $0 \leq \theta \leq \pi-\omega$,
but there are other restrictions.
The point $c$ must be strictly to the right of $v_j$,
so $\beta-\omega < \theta$.
Also, $b$ must be strictly to the left of $v_j$, 
so $\theta < \pi-\alpha$.
Finally,
$q$ must not reach the point
where the line through $q$ and $w$
is parallel to $R''$.
Hence $\theta < \arctan(\lambda)-\omega$.
To summarize,
we have
\begin{align}
\label{lemma arc fixed midpoint on a line item constraint 1}
0 \leq &\,\theta \leq \pi-\omega \enspace,\\
\label{lemma arc fixed midpoint on a line item constraint 2}
\beta-\omega < &\,\theta < \min(\pi-\alpha,\arctan(\lambda)-\omega) \enspace.
\end{align}

By elementary trigonometry,
we get
$$q = (2r\cos(\theta)\sin(\theta+\omega),2r\sin(\theta)\sin(\theta+\omega)) \enspace.$$
Then,
finding the general equation of the line through $q$ and $v$,
and of the line through $q$ and $w$,
we can calculate the general coordinates of $b$ and $c$:
\begin{eqnarray*}
b &=& \left(\frac{\mu\cot(\theta)}{1-\lambda\cot(\theta)},\frac{\mu}{1-\lambda\cot(\theta)}\right) \enspace,\\
c &=& \left(\frac{2r\sin^2\!(\omega)\cot(\theta)+2r\sin(\omega)\cos(\omega)+\mu\cos(\omega)\cot(\theta)-\mu\sin(\omega)}
{\sin(\omega)\cot(\theta)+\cos(\omega)-\lambda\cos(\omega)\cot(\theta)+\lambda\sin(\omega)}\right.,\\
&&\phantom{(}\left.\frac{(2r\sin(\omega)\lambda+\mu)(\sin(\omega)\cot(\theta)+\cos(\omega))}
{\sin(\omega)\cot(\theta)+\cos(\omega)-\lambda\cos(\omega)\cot(\theta)+\lambda\sin(\omega)}\right) \enspace.
\end{eqnarray*}
Therefore we want to find $\theta$ such that
$v_j$ is the midpoint of $bc$,
which leads to the equation
$$\frac{1}{2}\left(\frac{\mu\cot(\theta)}{1-\lambda\cot(\theta)} + \frac{2r\sin^2\!(\omega)\cot(\theta)+2r\sin(\omega)\cos(\omega)+\mu\cos(\omega)\cot(\theta)-\mu\sin(\omega)}
{\sin(\omega)\cot(\theta)+\cos(\omega)-\lambda\cos(\omega)\cot(\theta)+\lambda\sin(\omega)}\right) = s$$
or
\begin{eqnarray}
\nonumber
&&\phantom{+}(2r\sin^2\!(\omega)\lambda+2\mu\cos(\omega)\lambda-\mu\sin(\omega)-2s\lambda\sin(\omega)+2s\lambda^2\cos(\omega))\cot^2\!(\theta)\\
\nonumber
&&+(2r\sin(\omega)\cos(\omega)\lambda-2r\sin^2\!(\omega)-2\mu\cos(\omega)-2\mu\sin(\omega)\lambda+2s\sin(\omega)-4s\cos(\omega)\lambda\\
\nonumber
&&\phantom{(}-2s\lambda^2\sin(\omega))\cot(\theta)\\
\label{lemma arc fixed midpoint on a line equation}
&&+(\mu\sin(\omega)-2r\sin(\omega)\cos(\omega)+2s\cos(\omega)+2s\lambda\sin(\omega)) = 0 \enspace.
\end{eqnarray}
Since this is a quadratic equation in $\cot(\theta)$,
it is solvable in constant time 
for $\theta$ satisfying (\ref{lemma arc fixed midpoint on a line item constraint 1})
and (\ref{lemma arc fixed midpoint on a line item constraint 2})
(or it is possible to decide 
that there is no such solution in constant time).

\item[(2)] The case where $R''$ has non-negative slope
is similar to the case where $R''$ has strictly negative slope.

\item[(3)] Suppose $R''$ is vertical.
Therefore the general equation for $R''$ is $x=s$.
We must have
$0< \omega <\frac{1}{2}\pi$,
otherwise there is no solution.
There are three subcases to consider:
(3.1) $s\leq 0$,
(3.2) $s\in\,]0,2r\sin(\omega)[$
(3.3) or $s\geq 2r\sin(\omega)$.

\begin{enumerate}
\item[(3.1)] Suppose $R''$ is vertical and $s\leq 0$.
Trivially $0 \leq \theta \leq \pi-\omega$,
but there are other restrictions.
With the notation of the previous cases,
$w$ must be between $q$ and $c$,
therefore $\theta < \frac{1}{2}\pi-\omega$.
The point $b$ must be strictly above $v_j$
so $\theta < \pi-\alpha$.
Also,
$c$ must be strictly under $v_j$,
so $\beta-\omega < \theta$.
It all sums up to
\begin{align}
\label{lemma arc fixed midpoint on a line item constraint 5}
0 &\,\leq \theta \enspace,\\
\label{lemma arc fixed midpoint on a line item constraint 6}
\beta-\omega &\,< \theta < \min\left(\pi-\alpha,\frac{1}{2}\pi-\omega\right) \enspace.
\end{align}

By elementary trigonometry,
we get
$$q = (2r\cos(\theta)\sin(\theta+\omega),2r\sin(\theta)\sin(\theta+\omega)) \enspace.$$
Then,
finding the general equation of the line through $q$ and $v$,
and of the line through $q$ and $w$,
we can calculate the general coordinates of $b$ and $c$:
\begin{eqnarray*}
b &=& \left(s,\frac{s}{\cot(\theta)}\right) \enspace,\\
c &=& \left(s,\frac{(s-2r\sin(\omega))(\sin(\omega)\cot(\theta)+\cos(\omega))}{\cot(\theta)\cos(\omega)-\sin(\omega)}\right) \enspace.
\end{eqnarray*}
Therefore we want to find $\theta$ such that
$v_j$ is the midpoint of $bc$,
which leads to the equation
$$\frac{1}{2}\left(\frac{s}{\cot(\theta)} + \frac{(s-2r\sin(\omega))(\sin(\omega)\cot(\theta)+\cos(\omega))}{\cot(\theta)\cos(\omega)-\sin(\omega)}\right) = t$$
or
\begin{eqnarray}
\nonumber
&&\phantom{+}(2r\sin^2\!(\omega)+t\cos(\omega)-s\sin(\omega))\cot^2\!(\theta)\\
\nonumber
&&+(2r\sin(\omega)\cos(\omega)-2s\cos(\omega)-t\sin(\omega))\cot(\theta)\\
\label{eqn quad proof 2}
&&+s\sin(\omega) = 0 \enspace.
\end{eqnarray}
Since this is a quadratic equation in $\cot(\theta)$,
it is solvable in constant time 
for $\theta$ satisfying (\ref{lemma arc fixed midpoint on a line item constraint 5})
and (\ref{lemma arc fixed midpoint on a line item constraint 6})
(or it is possible to decide 
that there is no such solution in constant time).

\item[(3.2)] Suppose $R''$ is vertical and $s\in\,]0,2r\sin(\omega)[$.
In this case there is no solution.

\item[(3.3)] The case where $R''$ is vertical and $s\geq 2r\sin(\omega)$
is similar to the case where $R''$ is vertical and $s\leq 0$.
\qed
\end{enumerate}
\end{enumerate}

\begin{figure}
\centering
\includegraphics[scale=1]{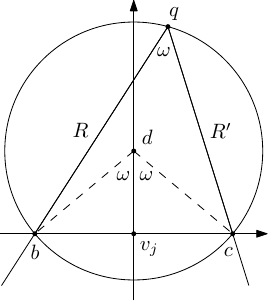}
\caption{Proof of Lemma~\ref{lemma vertex fixed midpoint on a segment}.\label{sommet-segment-point-proof}}
\end{figure}

\begin{lemma}
\label{lemma vertex fixed midpoint on a segment}
Let $q$ be a point,
$R''$ be a line
and $v_j \in R''$ be a point.
It is possible to find the triangle $\triangle qbc$
such that $\angle bqc=\omega$,
$b,c\in R''$
and $v_j$ is the midpoint of $bc$
in $O(1)$ time
(refer to Figure~\ref{sommet-segment-point-proof}).
\end{lemma}

\proof
Without loss of generality,
$R''$ is the $x$-axis,
$v_j=(0,0)$,
$q= \left(q_x,q_y\right)$ with $q_y>0$
and $b$ is to the left of $c$.
Denote by $d$
the center of the circumcircle of $\triangle qbc$.

Suppose $\omega\neq\frac{1}{2}\pi$.
By elementary geometry,
$d$ lies on the line segment bisector of $bc$
and $\angle bdv_j = \angle cdv_j = \omega$
(refer to Figure~\ref{sommet-segment-point-proof}).
Hence $d=(0,h)$ for an $h\in\mathbb{R}$.
Therefore,
the equation of the circumcircle is\footnote{We need
the assumption $\omega\neq\frac{1}{2}\pi$ here,
otherwise $\frac{h^2}{\cos^2\!(\omega)}$ is undefined.}
$$x^2+(y-h)^2 = \frac{h^2}{\cos^2\!(\omega)} \enspace.$$
Since $q$ is on this circle,
then
$$q_x^2+\left(q_y-h\right)^2 = \frac{h^2}{\cos^2\!(\omega)} \enspace,$$
so
$$h=\frac{\cos(\omega)}{\sin^2\!(\omega)}\left(-q_y\cos(\omega)+\sqrt{q_x^2\sin^2\!(\omega)+q_y^2}\right) \enspace,$$
from which
\begin{eqnarray*}
b &=& \left(q_y\cot(\omega)-\sqrt{q_x^2\sin^2\!(\omega)+q_y^2}\,\csc(\omega),0\right) \enspace,\\
c &=& \left(-q_y\cot(\omega)+\sqrt{q_x^2\sin^2\!(\omega)+q_y^2}\,\csc(\omega),0\right) \enspace.
\end{eqnarray*}

If $\omega=\frac{1}{2}\pi$,
then $d=(0,0)$ and the equation of the circumcircle is $x^2+y^2 = q_x^2+q_y^2$.
Therefore,
\begin{eqnarray*}
b &=& \left(-\sqrt{q_x^2+q_y^2},0\right) = \left(q_y\cot(\omega)-\sqrt{q_x^2\sin^2\!(\omega)+q_y^2}\,\csc(\omega),0\right) \enspace,\\
c &=& \left(\sqrt{q_x^2+q_y^2},0\right) = \left(-q_y\cot(\omega)+\sqrt{q_x^2\sin^2\!(\omega)+q_y^2}\,\csc(\omega),0\right) \enspace.
\end{eqnarray*}
\qed

There are $n'=O(n)$ event points of the first type.
There are at most $n$ event points of the second type
because there are $n$ edges.
For the same reason,
there are at most $n$ event points of the third type.
Therefore,
{\bf Step 3} 
(computing the event points)
takes $O(n)$ time.

\section{Finding the Optimal Solution when the Apex is on a Circular Arc}
\label{section optimal solution apex on circular arc}

Between any two consecutive event points,
there is a single arc of a circle
(that might be reduced to a single point
if one of the event points is on one of the vertices of $P$).
Consider such an arc.
As $q$ moves along it,
the minimum-area triangle enclosing $P$ changes.
In this section,
we show how to compute the optimal triangle
for a fixed arc of a circle
between two consecutive event points.
Four different cases can occur:
\begin{itemize}
\item either the circular arc is reduced to a single point or not.
If the former is true,
then the apex $q$ is forced to stay on one of the vertices of $P$.

\item Either the midpoint $m$ is forced to stay on a vertex of $P$ or not.
\end{itemize}
The following lemmas describe how to compute 
the minimum-area triangle enclosing $P$
in these four different situations.

\begin{figure}
\centering
\includegraphics[scale=1]{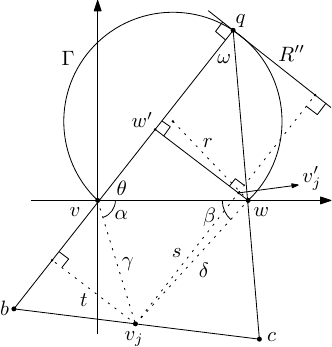}
\caption{Proof of Lemma~\ref{lemma arc fixed point minimal triangle}.\label{arc-circle-point-minima-proof}}
\end{figure}

Lemma~\ref{lemma arc fixed point minimal triangle}
describes how to compute the minimum-area enclosing triangle
when $q$ moves along an arc of a circle
and $m$ is forced to stay on one of the vertices of $P$.
\begin{lemma}
\label{lemma arc fixed point minimal triangle}
Let $\Gamma = [v,w]$ be an arc of a circle
and $v_j$ be a point.
It is possible to find the triangle $\triangle qbc$
of minimum area such that $q \in \Gamma$,
$q$, $v$ and $b$ lie on the same line,
$q$, $w$ and $c$ lie on the same line,
and $v_j \in bc$
in $O(1)$ time
(refer to Figure~\ref{arc-circle-point-minima-proof}).
\end{lemma}

\proof
The strategy is to first fix a point $q$ on $\Gamma$
and then follow the proof of Lemma~\ref{lemma fixed wedge minimal triangle}
in order to construct the minimum-area triangle for this fixed $q$.
Then we move $q$ along $\Gamma$
while maintaining the minimum-area triangle.
The smallest one among all these minimal triangles is optimal.

Without loss of generality,
$\Gamma$ is the locus of the point $q$
such that $\angle vqw = \omega$.
Hence we can take
$v=(0,0)$ and $w=(2r\sin(\omega),0)$
where $r$ is the radius of $\Gamma$.
Let $q$ be any point on $\Gamma$
(refer to Figure~\ref{arc-circle-point-minima-proof}).
Note $\gamma = |v_jv|$,
$\delta=|v_jw|$,
$\alpha = \angle v_jvw$
($0<\alpha<\pi$),
$\beta = \angle v_jwv$
($0<\beta<\pi$ and $0<\alpha+\beta<\pi$)
and $\theta = \angle qvw$.
Let $w'$ be the orthogonal projection of $w$
onto the line through $q$ and $v$.
Let $v_j'$ be the orthogonal projection of $v_j$
onto the line through $w$ and $w'$.
Let $R''$ be the tangent to $\Gamma$ at $q$.
Finally, 
note $s=dist(v_j,R'')$ and $t=|v_jv_j'|$.

By elementary trigonometry,
we get
$$q = (2r\cos(\theta)\sin(\theta+\omega),2r\sin(\theta)\sin(\theta+\omega)) \enspace.$$
By elementary trigonometry and geometry,
we get
\begin{eqnarray*}
s &=& |v_jv_j'| + |w'q|\\
&=& \delta\sin\left(\frac{1}{2}\pi-\theta+\beta\right) + 2r\sin(\omega)\sin(\theta)\cot(\omega)\\
&=& \delta\cos(\theta-\beta) + 2r\cos(\omega)\sin(\theta) \enspace,\\
t &=& \gamma\sin(\pi-\theta-\alpha)\\
&=& \gamma\sin(\theta+\alpha) \enspace.
\end{eqnarray*}

With respect to the proof of Lemma~\ref{lemma fixed wedge minimal triangle},
let $b$ (respectively $c$)
be the point on the line $qv$
(respectively on the line $qw$)
such that $|qb| = 2(s-t\cot(\omega))$
(respectively $|qc| = 2t\csc(\omega)$).
Therefore $\triangle qbc$ 
is of minimum area for this fixed $q$
(and therefore for this fixed $\theta$)
by Lemma~\ref{lemma fixed wedge minimal triangle}.

We need to study how the area of $\triangle qbc$ evolves 
as $q$ moves along $\Gamma$.
Trivially $0 \leq \theta \leq \pi-\omega$,
but there are other restrictions.
Since $v_j$ must not be on the line through $qv$,
$\theta < \pi-\alpha$.
Also,
$v_j$ must not be on the line through $qw$,
so $\pi-\beta < \theta$.
Finally,
we must have $|qv| \leq |qb|$ and $|qw| \leq |qc|$,
so $\arccot\left(\frac{1}{2}(\cot(\beta)-\cot(\alpha))\right)-\omega\leq\theta$
and $\theta\leq\arccot\left(\frac{1}{2}(\cot(\beta)-\cot(\alpha))\right)$.
It all sums up to
\begin{align}
\label{lemma arc fixed point minimal triangle constraint 1}
\max\left(0,\arccot\left(\frac{1}{2}(\cot(\beta)-\cot(\alpha))\right)-\omega\right) &\,\leq \theta \leq \min\left(\pi-\omega,\arccot\left(\frac{1}{2}(\cot(\beta)-\cot(\alpha))\right)\right) \enspace,\\
\label{lemma arc fixed point minimal triangle constraint 2}
\beta-\omega &\,< \theta < \pi-\alpha \enspace.
\end{align}
By the proof of Lemma~\ref{lemma fixed wedge minimal triangle},
the area of $\triangle qbc$ 
as a function of $\theta$ is
\begin{eqnarray*}
\sigma(\theta) &=& 2(s-t\cot(\omega))t\\
&=& 2((\delta\cos(\theta-\beta) + 2r\cos(\omega)\sin(\theta))-(\gamma\sin(\theta+\alpha))\cot(\omega))(\gamma\sin(\theta+\alpha))\\
&=& 2\gamma\sin(\theta+\alpha)(\delta\cos(\theta-\beta) + 2r\cos(\omega)\sin(\theta)-\gamma\cot(\omega)\sin(\theta+\alpha)) \enspace.
\end{eqnarray*}
With the sine law applied on $\triangle vv_jw$,
one gets
\begin{eqnarray*}
\gamma &=& \frac{2r\sin(\omega)\sin(\beta)}{\sin(\alpha+\beta)} \enspace,\\
\delta &=& \frac{2r\sin(\omega)\sin(\alpha)}{\sin(\alpha+\beta)} \enspace.
\end{eqnarray*}
Therefore,
\begin{eqnarray*}
\sigma(\theta) &=& 2\gamma\sin(\theta+\alpha)(\delta\cos(\theta-\beta) + 2r\cos(\omega)\sin(\theta)-\gamma\cot(\omega)\sin(\theta+\alpha))\\
&=& -\frac{8r^2\sin(\alpha)\sin(\beta)\sin(\omega)}{\sin^2\!(\alpha+\beta)}\sin(\beta-\omega-\theta)\sin(\alpha+\theta)\enspace.
\end{eqnarray*}

We have
\begin{eqnarray*}
\sigma'(\theta) &=& -\frac{8r^2\sin(\alpha)\sin(\beta)\sin(\omega)}{\sin^2\!(\alpha+\beta)}(\sin(\beta-\omega-\theta)\cos(\alpha+\theta) - \cos(\beta-\omega-\theta)\sin(\alpha+\theta))\\
&=& \frac{8r^2\sin(\alpha)\sin(\beta)\sin(\omega)}{\sin^2\!(\alpha+\beta)}\sin(\alpha-\beta+\omega+2\theta) \enspace.
\end{eqnarray*}
In order to find the candidates for minimum and maximum of $\sigma$,
we need to solve
\begin{eqnarray}
\label{lemma arc fixed point minimal triangle equation 1}
\sin(\alpha-\beta+\omega+2\theta) = 0
\end{eqnarray}
for $\theta$ satisfying (\ref{lemma arc fixed point minimal triangle constraint 1})
and (\ref{lemma arc fixed point minimal triangle constraint 2}).

We look for solutions to $\alpha-\beta+\omega+2\theta = k\pi$
for $k\in\mathbb{Z}$.
At first sight,
we have $\theta = \frac{k\pi-\alpha+\beta-\omega}{2}$,
but we need to study this general solution
in order for it to satisfy (\ref{lemma arc fixed point minimal triangle constraint 1})
and (\ref{lemma arc fixed point minimal triangle constraint 2}).
Putting this solution against these constraints,
we get
\begin{align}
\label{lemma arc fixed point minimal triangle constraint 3}
\max\left(0,\arccot\left(\frac{1}{2}(\cot(\beta)-\cot(\alpha))\right)-\omega\right) &\,\leq \frac{k\pi-\alpha+\beta-\omega}{2} \enspace,\\ 
\label{lemma arc fixed point minimal triangle constraint 4}
\frac{k\pi-\alpha+\beta-\omega}{2} &\,\leq \min\left(\pi-\omega,\arccot\left(\frac{1}{2}(\cot(\beta)-\cot(\alpha))\right)\right) \enspace,\\
\label{lemma arc fixed point minimal triangle constraint 5}
\beta-\omega &\,< \frac{k\pi-\alpha+\beta-\omega}{2} < \pi-\alpha \enspace,
\end{align}
which simplify to
\begin{align}
\label{lemma arc fixed point minimal triangle constraint 6}
\!\!\!\max\left(\pm\left(-k\pi+\alpha-\beta+2\,\arccot\left(\frac{1}{2}(\cot(\beta)-\cot(\alpha))\right)\right)\right) &\,\leq \omega \enspace,\\
\label{lemma arc fixed point minimal triangle constraint 7}
\omega &\,\leq \min\left(k\pi-\alpha+\beta,-(k-2)\pi+\alpha-\beta\right) \enspace,\\
\label{lemma arc fixed point minimal triangle constraint 8}
\max\left(-k\pi+\alpha+\beta,(k-2)\pi+\alpha+\beta\right) &\,< \omega \enspace.
\end{align}
If $k \leq 0$,
(\ref{lemma arc fixed point minimal triangle constraint 8}) and (\ref{lemma arc fixed point minimal triangle constraint 7})
lead to $-k\pi+\alpha+\beta < \omega \leq k\pi-\alpha+\beta < -k\pi+\alpha+\beta$,
which is a contradiction.
If $k \geq 2$,
(\ref{lemma arc fixed point minimal triangle constraint 8}) and (\ref{lemma arc fixed point minimal triangle constraint 7})
lead to $(k-2)\pi+\alpha+\beta < \omega \leq -(k-2)\pi+\alpha-\beta < (k-2)\pi+\alpha+\beta$,
which is a contradiction.
If $k=1$,
then $\theta = \frac{\pi-\alpha+\beta-\omega}{2}$
is a valid solution provided that
$$\max\left(\pm\left(\pi-\alpha+\beta-2\,\arccot\left(\frac{1}{2}(\cot(\beta)-\cot(\alpha))\right)\right)\right)\leq\omega\leq\min(\pi-\alpha+\beta,\pi+\alpha-\beta) \enspace.$$

In this case,
$\theta = \frac{\pi-\alpha+\beta-\omega}{2}$ is a maximum as we show below.
$$\sigma''(\theta) = \frac{16r^2\sin(\alpha)\sin(\beta)\sin(\omega)}{\sin^2\!(\alpha+\beta)}\cos(\alpha-\beta+\omega+2\theta)$$
and
\begin{eqnarray*}
\sigma''\left( \frac{\pi-\alpha+\beta-\omega}{2} \right) &=& \frac{16r^2\sin(\alpha)\sin(\beta)\sin(\omega)\cos(\pi)}{\sin^2\!(\alpha+\beta)} \\
&=&-\frac{16r^2\sin(\alpha)\sin(\beta)\sin(\omega)}{\sin^2\!(\alpha+\beta)}\\
&<& 0
\end{eqnarray*}
so $\theta = \frac{\pi-\alpha+\beta-\omega}{2}$ is a maximum.

Otherwise,
(\ref{lemma arc fixed point minimal triangle equation 1}) has no solution,
therefore $\sigma(\theta)$ is monotonic.
The conclusion is that
the minimum of $\sigma$
is at one (or both) of the extremities
of the domain of $\sigma$.
\qed

Looking at the proof of Lemma~\ref{lemma arc fixed point minimal triangle},
it suggests that the minimum-area triangle might not exist.
Indeed,
if $0 < \beta-\omega$
and $\pi-\alpha < \pi-\omega$,
then $\beta-\omega < \theta < \pi-\alpha$
by (\ref{lemma arc fixed point minimal triangle constraint 1})
and (\ref{lemma arc fixed point minimal triangle constraint 2}).
But since the minimum of $\sigma$
is at one (or both) of the extremities
of the domain of $\sigma$,
then it does not exist.
However,
in the setting of the general problem,
this does not occur.
By Proposition~\ref{proposition n-gon wedge minimal triangle},
the minimum-area triangle always exists
for a given $\omega$-wedge
so $\theta$ will always vary inside an interval
that includes its extremities.

Lemma~\ref{lemma arc fixed line minimal triangle}
describes how to compute the minimum-area enclosing triangle
when $q$ moves along an arc of a circle
and $m$ is forced to stay on one of the edges of $P$.
\begin{lemma}
\label{lemma arc fixed line minimal triangle}
Let $\Gamma = [v,w]$ be an arc of a circle
and $R''$ be a line.
It is possible to find the point $q\in\Gamma$
such that the line through $qv$,
the line through $qw$
and $R''$
form a triangle of minimum area
in $O(1)$ time
(refer to Figure~\ref{arc-circle-segment-minima-proof}).
\begin{figure}
\centering
\includegraphics[scale=1]{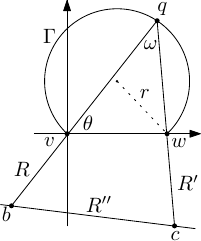}
\caption{Proof of Lemma~\ref{lemma arc fixed line minimal triangle}.\label{arc-circle-segment-minima-proof}}
\end{figure}
\end{lemma}

The proof of Lemma~\ref{lemma arc fixed line minimal triangle}
contains several technical details.
It is presented in~\ref{appendix proof lemma}.
Lemma~\ref{lemma vertex fixed point minimal triangle}
describes how to compute the minimum-area triangle enclosing $P$
when $q$ is forced to stay on one of the vertices of $P$
and $m$ is forced to stay on another one of the vertices of $P$.
\begin{lemma}
\label{lemma vertex fixed point minimal triangle}
Let $q$ and $v_j$ be two points.
It is possible to find the triangle $\triangle qbc$
of minimum area such that 
$v_j$ is the midpoint of $bc$
and $\angle bqc = \omega$
in $O(1)$ time.
\end{lemma}

\begin{figure}
\centering
\includegraphics[scale=1]{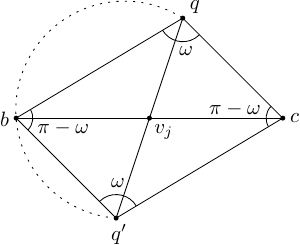}
\caption{Proof of Lemma~\ref{lemma vertex fixed point minimal triangle}.\label{sommet-point-minima-proof-2}}
\end{figure}

\proof Let $q'$ be a point such that $|qv_j| = |q'v_j|$
and $q$, $v_j$ and $q'$ are aligned
(refer to Figure~\ref{sommet-point-minima-proof-2}).
Let $b$ and $c$ be any points such that 
$|bv_j| = |cv_j|$ and $b$, $v_j$ and $c$ are aligned.
By construction,
the quadrilateral $[qcq'b]$ is a parallelogram,
$\angle bqc = \angle bq'c = \omega$,
$\angle qbq' = \angle qcq' = \pi-\omega$
and the areas of $\triangle qbc$ and $\triangle qbq'$ are equal.
Moreover,
$b$ is on a circular arc
that is the locus of the point $b$
such that $\angle qbq' = \pi-\omega$.

Since $|qq'|$ is fixed and the areas of $\triangle qbc$ and $\triangle qbq'$ are equal,
it suffices to minimize the height of $\triangle qbq'$ relative to $qq'$.
The solution is to take $b$ as close as possible to $qq'$.
\qed

Lemma~\ref{lemma vertex fixed line minimal triangle}
describes how to compute the minimum-area enclosing triangle
when $q$ is forced to stay on one of the vertices of $P$
and $m$ is forced to stay on one of the edges of $P$.
\begin{lemma}
\label{lemma vertex fixed line minimal triangle}
Let $q$ be a point
and $R''$ be a line.
The triangle $\triangle qbc$
of minimum area such that 
$b,c\in R''$
and $\angle bqc = \omega$
is isosceles
and can be found
in $O(1)$ time.
\end{lemma}

\proof
Let $\triangle qb'c'$ be any triangle such that $b',c'\in R''$,
$\angle bq'c' = \omega$
and $b'$ is to the left of $c'$.
We prove that the area of $\triangle qb'c'$ is bigger than or equal to the area of the isosceles triangle $\triangle qbc$.
Without loss of generality,
assume that $b'$ is to the left of $b$.

\begin{figure}
\centering
\includegraphics[scale=1]{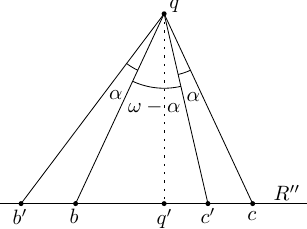}
\caption{Proof of Lemma~\ref{lemma vertex fixed line minimal triangle}.\label{sommet-droite-minima-proof-2}}
\end{figure}

Let $q'$ be the projection of $q$ onto $R''$.
From elementary geometry,
as a point $x\in R''$ gets closer to $q'$,
$|qx|$ decreases.
Conversely,
as $x$ gets farther from $q'$,
$|qx|$ increases.
Therefore,
since the area of $\triangle qb'c'$ is $\frac{1}{2}|qb'||qc'|\sin(\omega)$,
we may assume that $q'\in b'c'$
(refer to Figure~\ref{sommet-droite-minima-proof-2}).
Note $\alpha = \angle b'qb$.
Therefore,
$\angle c'qc = \alpha$
and $\angle bqc' = \omega-\alpha$.
Thus,
it is sufficient to prove that the area of $\triangle qb'b$
is bigger than or equal to the area of $\triangle qc'c$.
We have
$$\frac{1}{2}|qb'||qb|\sin(\alpha) \geq \frac{1}{2}|qb|^2\sin(\alpha) = \frac{1}{2}|qc|^2\sin(\alpha) \geq \frac{1}{2}|qc'||qc|\sin(\alpha) \enspace,$$
which completes the proof.
\qed

\begin{description}
\item[Step 4] For each pair of consecutive event points,
compute the minimum-area triangle enclosing $P$
and having a vertex on the corresponding arc of a circle.
\end{description}

Since there are $O(n)$ event points,
{\bf Step 4} takes $O(n)$ time.

\section{Putting it All Together}
\label{section putting all together}

The vertex that subtends the angle $\omega$ 
of the optimal triangle enclosing $P$
has to be on the $\omega$-cloud $\Omega$.
$\Omega$ was computed in {\bf Step 1}
in $O(n)$ time
(refer to Section~\ref{section overview preliminaries}).
In {\bf Step 2}
(refer to Section~\ref{section optimal solution given omega wedge}),
we fixed an $\omega$-wedge $W$
and computed the minimum-area triangle enclosing $P$
that can be constructed with $W$
in $O(n)$ time.
This triangle is such that
the midpoint $m$ of its third side is on $P$
(refer to Proposition~\ref{proposition n-gon wedge minimal triangle} 
and Corollary~\ref{corollary n-gon wedge minimal triangle}).
Then we divided $\Omega$ into a linear number of pieces
in {\bf Step 3}
(refer to Section~\ref{section walking around omega cloud}).
Within each of these pieces,
an optimal triangle can be computed in $O(1)$ time.
This was done in {\bf Step 4}
(refer to Section~\ref{section optimal solution apex on circular arc}).

\begin{description}
\item[Step 5] Find the smallest area triangles among those calculated 
in {\bf Step 4}.
\end{description}

Since there are $O(n)$ event points,
{\bf Step 5} takes $O(n)$ time.
Since each step takes no longer than $O(n)$ time,
the algorithm takes $O(n)$ time.
If the input is a set of points,
compute the convex hull
and then apply this algorithm.
In this situation,
the computation takes $O(n\log n)$ time
because of the computation of the convex hull.

We summarize the final algorithm.
\begin{algorithm}(Minimum-Area Triangle with a Fixed Angle Enclosing $S$)
\label{algorithm minimum enclosing triangle fixed angle}
\begin{itemize}
\item INPUT: A finite set $S$ of points in the plane
and an angle $0<\omega<\pi$.

\item OUTPUT: All triangles with minimum area 
having an angle of $\omega$
that enclose $S$.
\end{itemize}
\begin{enumerate}
\item[0.] Compute the convex hull of $S$ and denote it by $P$.
We denote the edges and the vertices of $P$
in clockwise order 
by $e_i$ and $v_i$ for $0\leq i \leq n-1$
(all index manipulation is modulo $n$).

\item Compute the $\omega$-cloud around $P$
and denote it by $\Omega$
(refer to Section~\ref{section overview preliminaries}).
$\Omega$ consists of $n'=O(n)$ circular arcs
that we denote in clockwise order by $\Gamma_i$ for $0\leq i \leq n'-1$.
The intersection point of $\Gamma_i$ and $\Gamma_{i+1}$ 
is denoted by $u_{i+1}$ for $0\leq i \leq n'-1$.

\item\label{step minimum enclosing triangle fixed omega-wedge}  
Let $q \in \Gamma_0$ be such that $q=u_0$.
Consider the $\omega$-wedge $W=\wedge(\omega,q,R,R')$ 
that touches $P$.
Apply Algorithm~\ref{algorithm minimum enclosing triangle fixed wedge}
with $P$ and $W$
(refer to Section~\ref{section optimal solution given omega wedge}).
Let $\triangle qbc$ be the output 
of Algorithm~\ref{algorithm minimum enclosing triangle fixed wedge}
and denote by $m$ the midpoint of segment $bc$.

\item\label{step event points} 
Move $q$ clockwise along $\Omega$
and maintain $W$, $b$, $c$ and $m$
as defined in~\ref{step minimum enclosing triangle fixed omega-wedge}.
Collect all of the following three types of event points
(see Section~\ref{section walking around omega cloud}
for formal definition):
\begin{description}
\item[Type 1]
$q$ is on the intersection point of two consecutive circular arcs of $\Omega$ 
for an $i$ with $0\leq i \leq n'-1$.

\item[Type 2]
$q$ is such that the third side $bc$ of the triangle
is on an edge $e_i$ of $P$
(for an $i$ with $0\leq i \leq n-1$)
and the midpoint $m$ of $bc$
is on the first vertex $v_i$ of $e_i$
(when the vertices of $P$ are considered in clockwise order).

\item[Type 3]
$q$ is such that the third side $bc$ of the triangle
is on an edge $e_i$ of $P$
(for an $i$ with $0\leq i \leq n-1$)
and the midpoint $m$ of $bc$
is on the last vertex $v_{i+1}$ of $e_i$
(when the vertices of $P$ are considered in clockwise order).
\end{description}

\item\label{step minimum enclosing triangle between two event points}
For each pair of consecutive event points
computed in~\ref{step event points},
which we denote by $(q',q'')$,
there is a single arc of a circle.
This circular arc 
might be reduced to a single point
if two consecutive event points 
are on one of the vertices of $P$.
When $q$ moves on such an arc,
either $m$ is forced to stay 
on one of the vertices of $P$
or $m$ is forced to stay on one of the edges of $P$
(see Sections~\ref{section overview preliminaries}
and~\ref{section walking around omega cloud}
for complete discussion).
For each pair $(q',q'')$ of consecutive event points,
\begin{itemize}
\item if the circular arc between $q'$ and $q''$ 
is not reduced to a single point
and if $m$ is forced to stay 
on one of the vertices of $P$,
store the triangle defined by Lemma~\ref{lemma arc fixed point minimal triangle}
(refer to Section~\ref{section optimal solution apex on circular arc}).

\item If the circular arc between $q'$ and $q''$ 
is not reduced to a single point
and if $m$ is forced to stay 
on one of the edges of $P$,
store the triangle defined by Lemma~\ref{lemma arc fixed line minimal triangle}
(refer to Section~\ref{section optimal solution apex on circular arc}).

\item If the circular arc between $q'$ and $q''$ 
is reduced to a single point
and if $m$ is forced to stay 
on one of the vertices of $P$,
store the triangle defined by Lemma~\ref{lemma vertex fixed point minimal triangle}
(refer to Section~\ref{section optimal solution apex on circular arc}).

\item If the circular arc between $q'$ and $q''$ 
is reduced to a single point
and if $m$ is forced to stay 
on one of the edges of $P$,
store the triangle defined by Lemma~\ref{lemma vertex fixed line minimal triangle}
(refer to Section~\ref{section optimal solution apex on circular arc}).
\end{itemize}

\item Return the smallest area triangles among those stored
in~\ref{step minimum enclosing triangle between two event points}.
\end{enumerate}
\end{algorithm}

\section{A Note on the Complexity of the Solution}
\label{section complexity solution}

Notice that one of the cases 
in Section~\ref{section optimal solution apex on circular arc} 
required us to find the roots of a fourth degree polynomial
(refer to Lemma~\ref{lemma arc fixed line minimal triangle}). 
One may ask whether or not 
this is necessary or if one can avoid 
finding the roots of such a polynomial
to solve this problem.
In this section, 
we show that it is unavoidable in certain situations, 
by providing an example of a set of points 
where the optimal solution lies on the root 
of an irreducible fourth degree polynomial.

\begin{figure}
\centering
\includegraphics[scale=1]{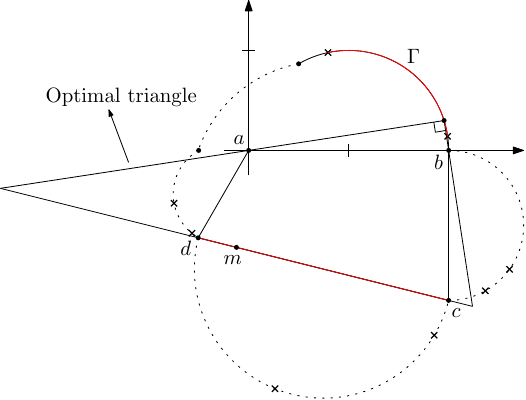}
\caption{Example where the optimal solution involves the roots of a quartic equation.\label{figure quartic}}
\end{figure}

Consider the quadrilateral $[abcd]$ where
$a=(0,0)$,
$b=(2,0)$,
$c=\left(2,-\frac{3}{2}\right)$
and $d=\left(-4\frac{4\sqrt{3}-1}{47},4\frac{\sqrt{3}-12}{47}\right)$,
and take $\omega=\frac{1}{2}\pi$
(hence we are looking for the minimum-area enclosing right triangle).
It turns out that the optimal right triangle
is such that the right angle is on $\Gamma$
(refer to Figure~\ref{figure quartic})
and the hypotenuse is on $cd$.
Therefore, 
we need to solve (\ref{lemma arc fixed line minimal triangle equation 2})
with $\mu:=-1$,
$\lambda:=-\frac{1}{4}$
and $r:=1$.
It leads to the following quartic equation.
$$p_4(X)=13X^4-92X^3+45X^2+12X-62=0$$
This equation admits two real solutions $X_1$ and $X_2$.
Note 
\begin{eqnarray*}
\alpha &=& \sqrt[3]{-1107+6\sqrt{34134}} \enspace,\\
\beta &=& \sqrt{1726-\frac{1105}{\alpha}+\frac{221}{3}\alpha} \enspace,
\end{eqnarray*}
then
\begin{eqnarray*}
X_1 &=& \frac{23}{13}+\frac{1}{26}\beta-\frac{1}{2}\sqrt{\frac{3452}{169}-\frac{17}{39}\alpha+\frac{85}{13\alpha}+\frac{136796}{169\beta}}\approx -0.761694 \enspace,\\
X_2 &=& \frac{23}{13}+\frac{1}{26}\beta+\frac{1}{2}\sqrt{\frac{3452}{169}-\frac{17}{39}\alpha+\frac{85}{13\alpha}+\frac{136796}{169\beta}}\approx 6.543373 \enspace.
\end{eqnarray*}
Solving $\cot(\theta)=X_1$
and $\cot(\theta)=X_2$,
then testing against $\sigma(\theta)$
(refer to the proof of Lemma~\ref{lemma arc fixed line minimal triangle}),
one finds that $\theta \approx \arccot(6.543373) \approx 0.048273\,\pi$
minimizes $\sigma(\theta)$.

Since $p_4(X)$ and its resolvent cubic
$$\rho(X)=X^3-\frac{45}{13}X^2+\frac{2120}{169}X+\frac{377816}{2197}$$
are irreducible over $\mathbb{Q}[X]$,
the algebraic expressions for the roots of $p_4$ must be written 
with square roots and cubic roots.
Moreover,
they cannot be simplified (see~\cite{dummit}).

Therefore, 
in general, 
we cannot avoid square roots
nor cubic roots in
the computation of the minimum-area enclosing triangle
with a fixed angle.

Finally,
we prove an $\Omega(n\log n)$ time lower bound for this problem
in the algebraic computation tree model.
We use a reduction to Max-Gap problem for points in the first quadrant
of the unit circle (see~\cite{DBLP:journals/algorithmica/LeeW86}).
Our proof is constructed upon the proof of Theorem 5 in~\cite{DBLP:journals/ijcga/BoseMSS11}.
We first need the following lemma.
\begin{lemma}
\label{lemma smallest triangle circle}
The smallest area triangle enclosing a circle is an equilateral triangle.
\end{lemma}

\proof Let $\triangle abc$ be a triangle
and let $r$ be the radius of its incircle.
Then the area of $\triangle abc$ is
$$r^2\left(\cot\left(\frac{\angle abc}{2}\right)+\cot\left(\frac{\angle bca}{2}\right)+\cot\left(\frac{\angle cab}{2}\right)\right) \enspace,$$
which is minimized when $\angle abc = \angle bca = \angle cab$.
\qed

\begin{proposition}
Given a set $S$ of $n$ points and an angle $\omega$,
computing the triangle of minimum area with angle $\omega$ 
that encloses $S$
requires $\Omega(n \log n)$ operations in the algebraic computation tree model.
\end{proposition}

\proof
As in the proof of Theorem 5 in~\cite{DBLP:journals/ijcga/BoseMSS11},
let $Z = \{z_1, z_2,...,z_n\}$ be an instance of the Max-Gap
problem for points in the first quadrant of the unit circle centered at the origin of
the coordinates system,
where $z_i = (x_i,y_i)$,
for $i = 1,2,... ,n$.
We also define the set of points $S=Z \cup Z' \cup Z'' \cup\{a,b,c\}$
as in~\cite{DBLP:journals/ijcga/BoseMSS11}.

Let $T$ be the equilateral triangle in the proof of Theorem 5 in~\cite{DBLP:journals/ijcga/BoseMSS11}.
We prove that $T$
is the minimum-area triangle that encloses $S$.
Then,
the result follows since
the triangle of minimum area with angle $\omega=\frac{1}{3}\pi$
that encloses $S$ must be $T$.

Consider the convex hull of $S$.
By construction,
the smallest distance from the origin to an edge is
the distance between the origin and the edge that represents the maximum gap.
Therefore,
the incircle of $T$ is also enclosed in the convex hull of $S$.
Let $T'$ be any other triangle enclosing $S$.
Therefore,
it encloses the convex hull of $S$ and hence,
the incircle of $T$.
By Lemma~\ref{lemma smallest triangle circle},
the area of $T'$ cannot be smaller than the area of $T$.
\qed

With a similar argument,
we can prove the following lemma.
\begin{lemma}
\label{lemma lower bound ORourke}
Given a set $S$ of $n$ points,
computing the triangle of minimum area that encloses $S$
requires $\Omega(n \log n)$ operations in the algebraic computation tree model.
\end{lemma}

In~\cite{DBLP:journals/jal/ORourkeAMB86},
O'Rourke et al. show how to compute in $\Theta(n)$ time
the minimum-area triangle enclosing a convex $n$-gon,
which is optimal.
If the input is a set $S$ of $n$ points,
then the following strategy is optimal
by Lemma~\ref{lemma lower bound ORourke}.
Compute the convex hull $CH(S)$ of $S$
and then apply O'Rourke et al.'s algorithm on $CH(S)$.

\section{Conclusion}
\label{section conclusion}

We have shown how to compute 
all triangles of minimum area 
with a fixed angle $0<\omega<\pi$ 
that enclose a point set.
It would be interesting to see 
if some of these techniques generalize 
to other settings or other optimization criteria.
For example,
finding the smallest area tetrahedron with a fixed solid angle
of a set of points in three dimensions.

\bibliographystyle{alpha}
\bibliography{MinimumEnclosingAreaTriangleFixedAngle}

\appendix

\section{Proof of Lemma~\ref{lemma arc fixed line minimal triangle}}
\label{appendix proof lemma}

In this section,
we prove Lemma~\ref{lemma arc fixed line minimal triangle}.

\proof
Without loss of generality,
$\Gamma$ is the locus of the point $q$
such that $\angle vqw = \omega$.
Hence we can take
$v=(0,0)$,
and $w=(2r\sin(\omega),0)$
where $r$ is the radius of $\Gamma$.
Let $q$ be any point on $\Gamma$.
There are three cases to consider:
(1) $R''$ has strictly negative slope,
(2) $R''$ has non-negative slope
or (3) $R''$ is vertical.

\begin{enumerate}
\item[(1)] Suppose $R''$ has strictly negative slope.
The general equation for $R''$ is $y=\lambda x+\mu$ ($\lambda<0$).
Note $\theta = \angle qvw$.
Also note $R$ 
(respectively $R'$)
the line through $q$ and $v$
(respectively through $q$ and $w$).
Note $b$ 
(respectively $c$)
the intersection point of $R$ and $R''$
(respectively of $R'$ and $R''$).

Trivially $0 \leq \theta \leq \pi-\omega$,
but there is one more restriction.
The point $v$ must be between $q$ and $b$,
hence $\theta < \arctan(\lambda)-\omega$.
It all sums up to
\begin{eqnarray}
\label{lemma arc fixed line minimal triangle constraint 1}
&&0 \leq \theta \leq \pi-\omega \enspace,\\
\label{lemma arc fixed line minimal triangle constraint 2}
&&\theta < \arctan(\lambda)-\omega \enspace.
\end{eqnarray}

By elementary trigonometry,
we get
\begin{eqnarray*}
q &=& (2r\cos(\theta)\sin(\theta+\omega),2r\sin(\theta)\sin(\theta+\omega))\\
  &=& \left(2r\frac{\cos(\omega)\cot(\theta)+\sin(\omega)\cot^2\!(\theta)}{\cot^2\!(\theta)+1},2r\frac{\cos(\omega)+\sin(\omega)\cot(\theta)}{\cot^2\!(\theta)+1}\right)\enspace.
\end{eqnarray*}
Then,
finding the general equation of the line through $q$ and $v$,
and of the line through $q$ and $w$,
we can calculate the general coordinates of $b$ and $c$:
\begin{eqnarray*}
b &=& \left(\frac{\mu\cot(\theta)}{1-\lambda\cot(\theta)},\frac{\mu}{1-\lambda\cot(\theta)}\right) \enspace,\\
c &=& \left(\frac{2r\sin^2\!(\omega)\cot(\theta)+2r\sin(\omega)\cos(\omega)+\mu\cos(\omega)\cot(\theta)-\mu\sin(\omega)}
{\sin(\omega)\cot(\theta)+\cos(\omega)-\lambda\cos(\omega)\cot(\theta)+\lambda\sin(\omega)}\right.,\\
&&\phantom{(}\left.\frac{(2r\sin(\omega)\lambda+\mu)(\sin(\omega)\cot(\theta)+\cos(\omega))}
{\sin(\omega)\cot(\theta)+\cos(\omega)-\lambda\cos(\omega)\cot(\theta)+\lambda\sin(\omega)}\right) \enspace.
\end{eqnarray*}
Let $X(\theta) = \cot(\theta)$
(or $X$ for short).
Therefore,
\begin{eqnarray*}
q &=& \left(2r\frac{\sin(\omega)X^2+\cos(\omega)X}{X^2+1},2r\frac{\sin(\omega)X+\cos(\omega)}{X^2+1}\right)\enspace,\\
b &=& \left(\frac{-\mu X}{\lambda X-1},\frac{-\mu}{\lambda X-1}\right) \enspace,\\
c &=& \left(\frac{(2r\sin^2\!(\omega)+\mu\cos(\omega))X+2r\sin(\omega)\cos(\omega)-\mu\sin(\omega)}
{(\sin(\omega)-\lambda\cos(\omega))X+\cos(\omega)+\lambda\sin(\omega)}\right.,\\
&&\phantom{(}\left.\frac{(2r\sin(\omega)\lambda+\mu)(\sin(\omega)X+\cos(\omega))}
{(\sin(\omega)-\lambda\cos(\omega))X+\cos(\omega)+\lambda\sin(\omega)}\right) \enspace.
\end{eqnarray*}
Here is the area $\sigma$ of $\triangle qbc$:
\begin{eqnarray*}
&& \sigma(X)\\
&=& \frac{1}{2}|qb||qc|\sin(\omega)\\
&=& \frac{1}{2}\sin(\omega)\sqrt{\frac{((2\lambda r\sin(\omega)+\mu)X^2+2r(\lambda\cos(\omega)-\sin(\omega))X+\mu-2r\cos(\omega))^2}{(\lambda X-1)^2(1+X^2)}}\times\\
&&\sqrt{\frac{((2\lambda r\sin(\omega)+\mu)X^2+2r(\lambda\cos(\omega)-\sin(\omega))X+\mu-2r\cos(\omega))^2}{((\lambda\cos(\omega)-\sin(\omega))X-\lambda\sin(\omega)-\cos(\omega))^2(1+X^2)}}\\
&=&\frac{1}{2}\sin(\omega)\frac{((2\lambda r\sin(\omega)+\mu)X^2+2r(\lambda\cos(\omega)-\sin(\omega))X+\mu-2r\cos(\omega))^2}{\left|(\lambda X-1)((\lambda\cos(\omega)-\sin(\omega))X-\lambda\sin(\omega)-\cos(\omega))\right|(1+X^2)}\\
&=&\frac{1}{2}\sin(\omega)\frac{((2\lambda r\sin(\omega)+\mu)X^2+2r(\lambda\cos(\omega)-\sin(\omega))X+\mu-2r\cos(\omega))^2}{(1-\lambda X)(\lambda\sin(\omega)+\cos(\omega)-(\lambda\cos(\omega)-\sin(\omega))X))(1+X^2)} \enspace.
\end{eqnarray*}
The reason why the absolute value $|\cdot|$ disappeared
is twofold.
Firstly,
by (\ref{lemma arc fixed line minimal triangle constraint 2}),
$\theta < \arctan(\lambda)-\omega < \arctan(\lambda) = \arccot\left(\frac{1}{\lambda}\right)$,
so $\lambda X=\lambda \cot(\theta) < 1$.
Secondly,
by (\ref{lemma arc fixed line minimal triangle constraint 2}),
$\theta < \arctan(\lambda)-\omega$,
therefore
\begin{eqnarray*}
\theta &<& \arctan(\lambda)-\omega \\
\theta +\omega &<& \arctan(\lambda) \\
\theta +\omega &<& \arccot\left(\frac{1}{\lambda}\right) \\
\cot(\theta +\omega) &>& \frac{1}{\lambda} \qquad\qquad\textrm{$0<\theta+\omega<\pi$ by (\ref{lemma arc fixed line minimal triangle constraint 1}) and the hypothesis,} \\
\frac{\cot(\omega)\cot(\theta)-1}{\cot(\theta)+\cot(\omega)} &>& \frac{1}{\lambda} \\
\frac{\lambda\cot(\omega)\cot(\theta)-\lambda}{\cot(\theta)+\cot(\omega)} &<& 1 \qquad\qquad\textrm{$\lambda<0$ by the hypothesis,}\\
\lambda\cot(\omega)\cot(\theta)-\lambda &<& \cot(\theta)+\cot(\omega) \qquad\qquad\textrm{$0<\theta+\omega<\pi$ by (\ref{lemma arc fixed line minimal triangle constraint 1}) and the hypothesis,}\\
\phantom{\lambda\cot(\omega)\cot(\theta)-\lambda} && \phantom{\cot(\theta)+\cot(\omega)}\qquad\qquad\textrm{so $\cot(\theta)+\cot(\omega)>0$,}\\
\lambda\cot(\omega)X-\lambda &<& X+\cot(\omega)\\
(\lambda\cot(\omega)-1)X &<& \lambda+\cot(\omega)
\end{eqnarray*}
\begin{eqnarray*}
(\lambda\cos(\omega)-\sin(\omega))X &<& \lambda\sin(\omega)+\cos(\omega) \qquad\qquad\textrm{$0<\omega<\pi$ by the hypothesis,}\\
\phantom{(\lambda\cos(\omega)-\sin(\omega))X} && \phantom{\lambda\sin(\omega)+\cos(\omega)}\qquad\qquad\textrm{so $\sin(\omega)>0$.}
\end{eqnarray*}

Now we need to find for what values
of $X$ is $\sigma(X)$ minimum,
which means that we need to find for what $X$
does $\sigma'(X)=0$.
\begin{eqnarray*}
&&\sigma'(X)\\
&=&-\frac{1}{2}\sin(\omega)\left((2\lambda r\sin(\omega)+\mu )X^2+2r(\lambda \cos(\omega)-\sin(\omega))X+\mu -2r\cos(\omega)\right)\times\\
&&\left((-\mu \sin(\omega)+2r\sin^2\!(\omega)\lambda +2\lambda^3r\sin^2\!(\omega)+2\lambda \mu \cos(\omega)+\mu \lambda^2\sin(\omega) \right.\\
&&\phantom{((}+4\lambda^3r\cos^2\!(\omega)-4r\lambda^2\cos(\omega)\sin(\omega)) X^4\\
&&\phantom{(}+(-2r\sin^2\!(\omega)-2\mu \cos(\omega)-6\lambda^3r\sin(\omega)\cos(\omega)+2\lambda^2r\sin^2\!(\omega)\\
&&\phantom{(+(}+2\mu \lambda^2\cos(\omega)-4\mu \lambda \sin(\omega)-12\lambda^2r\cos^2\!(\omega)+10\lambda r\sin(\omega)\cos(\omega))X^3\\
&&\phantom{(}+6r(\cos(\omega)+\lambda \sin(\omega))(-\sin(\omega)+2\lambda \cos(\omega)+\lambda^2\sin(\omega))X^2\\
&&\phantom{(}+(2r\sin^2\!(\omega)-4r\cos^2\!(\omega)-2\mu \cos(\omega)-10\lambda^2r\sin^2\!(\omega)-4\mu \lambda \sin(\omega)\\
&&\phantom{(+(}+2\mu \lambda^2\cos(\omega)+2\lambda^3r\sin(\omega)\cos(\omega)-14\lambda r\sin(\omega)\cos(\omega))X\\
&&\phantom{(}+(4r\sin^2\!(\omega)\lambda -\mu \lambda^2\sin(\omega)-2\lambda \mu \cos(\omega)+2r\sin(\omega)\cos(\omega)\\
&&\phantom{(+(}\left.-2r\lambda^2\cos(\omega)\sin(\omega)+\mu \sin(\omega))\right)\div\\
&&\left((1-\lambda X)^2(-\lambda \sin(\omega)-\cos(\omega)+X\lambda \cos(\omega)-X\sin(\omega))^2(1+X^2)^2\right)
\end{eqnarray*}
Therefore $\sigma'(X)=0$ if and only if
\begin{eqnarray}
\label{lemma arc fixed line minimal triangle equation 1}
(2\lambda r\sin(\omega)+\mu)X^2+2r(\lambda \cos(\omega)-\sin(\omega))X+\mu -2r\cos(\omega)=0
\end{eqnarray}
or
\begin{eqnarray}
\nonumber
&&(-\mu \sin(\omega)+2r\sin^2\!(\omega)\lambda +2\lambda^3r\sin^2\!(\omega)+2\lambda \mu \cos(\omega)+\mu \lambda^2\sin(\omega)\\
\nonumber
&&\phantom{(}+4\lambda^3r\cos^2\!(\omega)-4r\lambda^2\cos(\omega)\sin(\omega)) X^4\\
\nonumber
&&+(-2r\sin^2\!(\omega)-2\mu \cos(\omega)-6\lambda^3r\sin(\omega)\cos(\omega)+2\lambda^2r\sin^2\!(\omega)\\
\nonumber
&&\phantom{+(}+2\mu \lambda^2\cos(\omega)-4\mu \lambda \sin(\omega)-12\lambda^2r\cos^2\!(\omega)+10\lambda r\sin(\omega)\cos(\omega))X^3\\
\nonumber
&&+6r(\cos(\omega)+\lambda \sin(\omega))(-\sin(\omega)+2\lambda \cos(\omega)+\lambda^2\sin(\omega))X^2\\
\nonumber
&&+(2r\sin^2\!(\omega)-4r\cos^2\!(\omega)-2\mu \cos(\omega)-10\lambda^2r\sin^2\!(\omega)-4\mu \lambda \sin(\omega)\\
\nonumber
&&\phantom{+(}+2\mu \lambda^2\cos(\omega)+2\lambda^3r\sin(\omega)\cos(\omega)-14\lambda r\sin(\omega)\cos(\omega))X\\
\nonumber
&&+(4r\sin^2\!(\omega)\lambda -\mu \lambda^2\sin(\omega)-2\lambda \mu \cos(\omega)+2r\sin(\omega)\cos(\omega)\\
\label{lemma arc fixed line minimal triangle equation 2}
&&\phantom{+(}-2r\lambda^2\cos(\omega)\sin(\omega)+\mu \sin(\omega))=0 \enspace.
\end{eqnarray}
The first equation is quadratic in $X=\cot(\theta)$,
therefore it can be solved in constant time for $X$
and then it remains to solve for $\theta$.
The second equation is quartic in $X=\cot(\theta)$,
therefore it can be solved\footnote{By Galois theory,
polynomials of degree $d\leq 4$ can be solved exactly in constant time.
Refer to~\cite{dummit}} 
in constant time for $X$
and then it remains to solve for $\theta$.

Actually,
what we are interested in is $\sigma'(X(\theta))=0$.
However,
$\sigma'(X(\theta))=\sigma'(X)X'(\theta)=\sigma'(X)\csc^2\!(\theta)$
and $\csc^2\!(\theta) \neq 0$.
Finally,
the extremities of the domain of $\theta$
are also candidates.

Overall, 
we get at most eight candidates for the minimum of $\sigma$
(at most two from the quadratic equation, 
at most four from the quartic equation
and the extremities of the domain of $\theta$).
So it can be solved exactly in constant time 
by taking the smallest of the eight candidates.

\item[(2)] The case where $R''$ has non-negative slope
is similar to the case where $R''$ has strictly negative slope.

\item[(3)] Suppose $R''$ is vertical.
The general equation of $ R$ is $x=s$
for $s\in\mathbb{R}$.
$0< \omega <\frac{1}{2}\pi$
otherwise there is no solution.
Using the notation of the previous cases,
there are three subcases to consider:
(3.1) $s\leq 0$,
(3.2) $s\in\,]0,2r\sin(\omega)[$
or (3.3) $s\geq 2r\sin(\omega)$.

\begin{enumerate}
\item[(3.1)] Suppose $R''$ is vertical and $s\leq 0$.
Trivially $0 \leq \theta \leq \pi-\omega$,
but there is one more restriction.
The point $v$ must be between $q$ and $b$,
hence $\theta < \frac{1}{2}\pi-\omega$.
It all sums up to
\begin{eqnarray}
\label{lemma arc fixed line minimal triangle constraint 7}
0 &\leq& \theta \enspace,\\
\label{lemma arc fixed line minimal triangle constraint 8}
\theta &<& \frac{1}{2}\pi-\omega \enspace.
\end{eqnarray}

By elementary trigonometry,
we get
\begin{eqnarray*}
q &=& (2r\cos(\theta)\sin(\theta+\omega),2r\sin(\theta)\sin(\theta+\omega))\\
  &=& \left(2r\frac{\cos(\omega)\cot(\theta)+\sin(\omega)\cot^2\!(\theta)}{\cot^2\!(\theta)+1},2r\frac{\cos(\omega)+\sin(\omega)\cot(\theta)}{\cot^2\!(\theta)+1}\right)\enspace.
\end{eqnarray*}
Then,
finding the general equation of the line through $q$ and $v$,
and of the line through $q$ and $w$,
we can calculate the general coordinates of $b$ and $c$:
\begin{eqnarray*}
b &=& \left(s,\frac{s}{\cot(\theta)}\right) \enspace,\\
c &=& \left(s,\frac{(s-2r\sin(\omega))(\sin(\omega)\cot(\theta)+\cos(\omega))}{\cos(\omega)\cot(\theta)-\sin(\omega)}\right) \enspace.
\end{eqnarray*}
We note $X(\theta)=\cot(\theta)$
(or $X$ for short).
Therefore,
\begin{eqnarray*}
q &=& \left(2r\frac{\sin(\omega)X^2+\cos(\omega)X}{X^2+1},2r\frac{\sin(\omega)X+\cos(\omega)}{X^2+1}\right)\enspace,\\
b &=& \left(s,\frac{s}{X}\right) \enspace,\\
c &=& \left(s,\frac{(s-2r\sin(\omega))(\sin(\omega)X+\cos(\omega))}{\cos(\omega)X-\sin(\omega)}\right) \enspace.
\end{eqnarray*}
Here is the area $\sigma$ of $\triangle qbc$.
\begin{eqnarray*}
&& \sigma(X)\\
&=& \frac{1}{2}|qb||qc|\sin(\omega)\\
&=& \frac{1}{2}\sin(\omega)
\sqrt{\frac{((2r\sin(\omega)-s)X^2+2r\cos(\omega)X-s)^2}{X^2(X^2+1)}}
\sqrt{\frac{((2r\sin(\omega)-s)X^2+2r\cos(\omega)X-s)^2}{(\cos(\omega)X-\sin(\omega))^2(X^2+1)}}\\
&=&\frac{1}{2}\sin(\omega)\frac{((2r\sin(\omega)-s)X^2+2r\cos(\omega)X-s)^2}{\left|X(\cos(\omega)X-\sin(\omega))\right|(X^2+1)}
\end{eqnarray*}
\begin{eqnarray*}
&=&\frac{1}{2}\sin(\omega)\frac{((2r\sin(\omega)-s)X^2+2r\cos(\omega)X-s)^2}{X(\cos(\omega)X-\sin(\omega))(X^2+1)}
\end{eqnarray*}
The reason why the absolute value $|\cdot|$ disappeared
is twofold.
Firstly,
by the hypothesis,
$0<\theta<\frac{1}{2}\pi$,
so $X=\cot(\theta)>0$.
Secondly,
by (\ref{lemma arc fixed line minimal triangle constraint 8}),
$\theta < \frac{1}{2}\pi-\omega$,
therefore
\begin{eqnarray*}
\theta &<& \frac{1}{2}\pi-\omega \\
\cot(\theta) &>& \cot\left(\frac{1}{2}\pi-\omega\right)\\
\cot(\theta) &>& \tan(\omega)\\
\cos(\omega)\cot(\theta) &>& \sin(\omega)\\
\cos(\omega)\cot(\theta) - \sin(\omega) &>& 0\\
\cos(\omega)X - \sin(\omega) &>& 0 \enspace.
\end{eqnarray*}

Now we need to find for what values
of $X$ where $\sigma(X)$ is minimum,
which means that we need to find for which $X$
does $\sigma'(X)=0$.
\begin{eqnarray*}
&&\sigma'(X)\\
&=&-\frac{1}{2}\sin(\omega)\left((2r\sin(\omega)-s)X^2+2r\cos(\omega)X-s\right)\times\\
&&\left((-s\sin(\omega)+2r\sin^2\!(\omega)+4r\cos^2\!(\omega))X^4-2\cos(\omega)(s+3r\sin(\omega))X^3 \right.\\
&&\phantom{(}\left. +6r\sin^2\!(\omega)X^2+2\cos(\omega)(r\sin(\omega)-s)X+s\sin(\omega)\right)\div\\
&&\left(X^2(\cos(\omega)X-\sin(\omega))^2(X^2+1)^2\right)
\end{eqnarray*}
Therefore $\sigma'(X)=0$ if and only if
\begin{eqnarray}
\label{lemma arc fixed line minimal triangle equation 3}
(2r\sin(\omega)-s)X^2+2r\cos(\omega)X-s=0
\end{eqnarray}
or
\begin{eqnarray}
\nonumber
&&(-s\sin(\omega)+2r\sin^2\!(\omega)+4r\cos^2\!(\omega))X^4-2\cos(\omega)(s+3r\sin(\omega))X^3 \\
\label{lemma arc fixed line minimal triangle equation 4}
&&+6r\sin^2\!(\omega)X^2+2\cos(\omega)(r\sin(\omega)-s)X+s\sin(\omega)=0 \enspace.
\end{eqnarray}
The first equation is quadratic in $X=\cot(\theta)$,
therefore it can be solved in constant time for $X$
and then it remains to solve for $\theta$.
The second equation is quartic in $X=\cot(\theta)$,
therefore it can be solved
in constant time for $X$
and then it remains to solve for $\theta$.

Actually,
what we are interested in is $\sigma'(X(\theta))=0$.
However,
$\sigma'(X(\theta))=\sigma'(X)X'(\theta)=\sigma'(X)\csc^2\!(\theta)$
and $\csc^2\!(\theta) \neq 0$.
Finally,
the extremities of the domain of $\theta$
are also candidates.

Overall, 
we get at most eight candidates for the minimum of $\sigma$
(at most two from the quadratic equation, 
at most four from the quartic equation
and the extremities of the domain of $\theta$).
So it can be solved exactly in constant time 
by taking the smallest of the eight candidates.

\item[(3.2)] Suppose $R''$ is vertical and $s\in\,]0,2r\sin(\omega)[$.
In this case there is no solution.

\item[(3.3)] The case where $R''$ is vertical and $s\geq 2r\sin(\omega)$
is similar to the case where $R''$ is vertical and $s\leq 0$.
\qed
\end{enumerate}
\end{enumerate}

\end{document}